\documentclass[english,11pt]{article}
\usepackage{multicol}
\usepackage{amsmath}
\usepackage{bbold}
\usepackage{mathrsfs}
\usepackage{graphicx}
\usepackage[table]{xcolor}
\usepackage{accents}
\usepackage{float}
\usepackage{color}
\RequirePackage[numbers,sort&compress]{natbib}
\usepackage[colorlinks=true,urlcolor=blue,linkcolor=blue,citecolor=red]{hyperref}

\def\sr{{\text{\tiny)}}}
\def\sl{{\text{\tiny(}}}
\def\ar{{\text{\tiny]}}}
\def\al{{\text{\tiny[}}}
\def\n{{\text{\tiny 0}}}
\def\on{{\text{\tiny 1}}}
\def\tw{{\text{\tiny 2}}}
\def\tr{{\text{\tiny 3}}}
\def\h{{\text{\tiny h}}}

\begin{document}
\begin{center}
{\Large \bf On black holes in new general relativity \\ }
\vskip 0.7cm
{D. F. L\'opez\footnote{diego.lopez@dal.ca} and A. A. Coley\footnote{alan.coley@dal.ca}}
\vskip 0.2cm
{\it Department of Mathematics and Statistics\\ 
Dalhousie University\\
Halifax, Canada}
\vskip 0.7cm
{R. J. van den Hoogen\footnote{rvandenh@stfx.ca}}
\vskip 0.2cm
{\it Department of Mathematics and Statistics\\ 
St. Francis Xavier University\\
Antigonish, Canada}\\
\vskip 1cm
\vskip 0.8cm
\begin{quote}
{\bf Abstract.}~{\small New General Relativity (NGR) is a class of teleparallel theories defined by three free parameters, effectively reduced to two after appropriate normalization, which are subject to experimental constraints. In this framework, matter couples minimally to the metric, ensuring that test particles follow geodesics and that null congruence expansions can be employed to detect local horizons. Assuming such horizons exist, we demonstrate that all physically viable NGR models—including the Teleparallel Equivalent of General Relativity (TEGR) and the one-parameter Hayashi and Shirafuji model (1P-H\&S)—inevitably exhibit divergences in torsion scalars at the local horizon. This singular behavior obstructs the interpretation of these models and their associated teleparallel geometries as black hole configurations.
}
\end{quote}
\end{center}
\vskip 1.0cm

%
%
\section{Introduction}

In teleparallel gravity (TG), the primary geometric object is torsion, which is derived from the frame (or co-frame) and a flat, metric-compatible spin connection \cite{aldrovandi1732013}. Understanding this geometric framework within which TG theories are formulated as alternatives to general relativity (GR) for describing gravitational interaction requires the study of spherically symmetric geometries. These geometries are essential for assessing the viability of gravitational theories. 
They encompass both time-dependent configurations, which are relevant for cosmological models that explain the large-scale evolution of the universe (as in GR through the Friedmann–Robertson–Walker (FRW) solution), and time-independent configurations that describe static black holes (such as the Schwarzschild solution in GR), which is a vacuum solution of GR.
However, such a solution raises philosophical issues in the teleparallel context: while singularities in GR typically occur only at the origin of the radial coordinate, from the literature we observe that TG consistently exhibits singular behaviour at both the origin and the putative horizon.

Within the family of TG theories, one stands apart due to its local dynamical equivalence with GR: the Teleparallel Equivalent of General Relativity (TEGR). This subclass is based on the torsion scalar \( T \), which appears in the gravitational action. 
Such an equivalence with GR is reflected in the equality of the field equations (FE) and hence the agreement of classical predictions \cite{aldrovandi1732013}, including the existence of an analogous Schwarzschild solution. 
However, the two theories differ conceptually and geometrically, which affects their  descriptions of black holes. In particular, while black holes are well-defined in GR, TEGR lacks a clear formulation of black holes.
It has been shown that the TEGR geometry features unavoidable divergences at the local horizon (defined below), indicating that the theory cannot consistently describe black holes~\cite{alanND2024}. For instance, in our analysis of static spherically symmetric vacuum spacetimes, we found that in the Schwarzschild-like solution of TEGR---unlike its GR counterpart---two  torsion invariants, diverge at the so-called Schwarzschild horizon. Moreover, the torsion scalar \( T \) also diverges at that location for certain choices of two free functions in the tetrad (when using a proper frame) or in the spin connection (when using the orthonormal frame). Consequently, in such cases within the TEGR, the horizon and the region inside it do not form part of the manifold. 

An important TG theory is $F(T)$ TG, which is a generalization of TEGR, where $F$ is an arbitrary twice-differentiable function of the torsion scalar $T$, appearing in the action \cite{bahamonde862023, cai792016, ferraro752007, krssak362019}. When the geometrical framework describing $F(T)$ TG is defined in a gauge invariant manner, the resulting FE derived from such a theory are fully Lorentz covariant \cite{krssak362019,krssak115}, and there will exist a frame/spin connection pair in which the spin connection identically vanishes known as the ``proper'' frame \cite{aldrovandi1732013, krssak362019}. Concerning static spherically symmetric vacuum solutions, a general family of such solutions for \( F(T) \) was obtained in \cite{pfeifer2021} by analyzing the effects of weak teleparallel perturbations on GR solutions. In \cite{coley616439}, the frame/spin connection pair describing the most general spherically symmetric teleparallel geometries was presented, and it was subsequently used in \cite{coley842024} to discover new solutions in \( F(T) \) TG, revealing that Birkhoff's theorem does not generally hold in \( F(T) \) theories. Static spherically symmetric solutions in $F(T)$ theory were reviewed in \cite{bahamonde862023}.  Finding exact vacuum static spherially symmetric $F(T)$ TG solutions is an open problem, and there are no known exact vacuum black hole solutions, aside from those analogous to the Schwarzschild solution. An exact non-trivial vacuum solution in $F(T)$ TG is the non-black hole power law solution presented in \cite{golov1}. A number of special static spherically symmetric non-vacuum solutions are known \cite{bahamonde862023,awad1,baha6,baha4}. Recently, it was shown that if a local horizon exists in $F(T)$ TG, the geometry is necessarily singular there. Hence, no black holes can exist in vacuum in the static spherically symmetric case \cite{alanND2024}. The $F(T)$ approach to axially symmetric vacuum solutions yields new solutions beyond the Kerr solution; however, these solutions still exhibit a singular horizon \cite{jarv2019flat,bahamonde2021exploring}.

Another notable TG theory is the New General Relativity (NGR), the straightforward modification of TEGR defined by incorporating additional torsion scalars, specifically the scalars constructed from the irreducible parts of the torsion ($\mathscr{V}$, $\mathscr{A}$, $\mathscr{T}$), via three free parameters ($c_{v}$, $c_{a}$, $c_{t}$) in its action. Various combinations of these scalars and parameters can be used, and TEGR emerges as a special case, in which the torsion scalar $T$ remains after all the parameters are fixed to particular values \cite{aldrovandi1732013}.
The NGR geometrical framework and its role as a gauge theory of the translational group were analysed in \cite{hayashi1967, si-ming1990}. This original approach, which involved a pure-gauge and Lorentz-restricted formulation, resulted in non-local (global) Lorentz-invariant FE. Such a conception of NGR, combined with the difficulty in determining symmetries in teleparallel geometries, produced neither general results nor well-behaved geometrical structures.  The viability of the theory was examined in \cite{kopczynski1982, hoissen1983, fukui1985, nester1988}, highlighting concerns regarding Lorentz invariance and the unpredictability of torsion behaviour. The local Lorentz invariance issue was resolved by extending the framework to include a non-vanishing spin connection alongside the tetrad, allowing for an invariant treatment of torsion \cite{krssak362019}. 

The phenomenological viability of NGR depends on a set of constraints that prevent the emergence of ghostly modes \cite{them} (see Table~\ref{GFC}). These constraints define a set of nine types of theory. The one-parameter NGR (1P-NGR) model, which falls under ``Type 2," has often been favored over other cases owing to its claimed absence of ghosts \citep{jimenez2020}. Although NGR may exhibit issues, particularly at the quantum level, it remains a justifiable framework for classical phenomenology, specifically in cosmological applications \citep{us}. In fact, the most general NGR model, referred to as ``Type 1," is likely to be healthy and may represent the most promising option, because it possesses a well-defined number of degrees of freedom and is the most robust in terms of physical modes \citep{gol}.

NGR cosmological models within a Lorentz-restricted framework were studied in \cite{fukuiS1984, mikhail1995, nester2008}. Kasner-type, Bianchi Type-I, and spatially homogeneous isotropic universe models were examined in \cite{fukuiS1984}. A comparison between a 2P-NGR family defined by \( c_{t} = 0 \) and 1P-NGR cosmological models was presented in \cite{mikhail1995}. A definition of ``gravitational energy" for spatially homogeneous cosmologies covering all nine Bianchi types was analyzed in \cite{nester2008}. More recently, a study on the invariant framework presented in \cite{golovnev2024} investigated conformal transformations in the most general parity-preserving NGR models and applied them to the analysis of cosmological perturbations in spatially flat cosmologies. The study concluded that 1P-NGR does not exhibit observable deviations from GR at that level of analysis.

Static spherically symmetric geometries in NGR were studied in \cite{hayashi191981,wanxiu1988,kawai831990,chengmin1995,obukhov672003, golovnev412024,asukula2311}. In its original Lorentz-restricted formulation \cite{hayashi191981}, NGR admits a spherically symmetric exact vacuum solution for the 3P family of the theory. However, this solution has been shown to describe a geometry with singularities at both the horizon and origin \cite{wanxiu1988,kawai831990,chengmin1995}. Using a fully invariant approach, it was further demonstrated that more general spherically symmetric solutions in NGR yield torsion invariants that still diverge at both the horizon and origin \cite{obukhov672003}. We review these solutions in Appendix~\ref{app}. Modern approaches have confirmed these findings and have uncovered new classes of solutions, although the issue of singularities at the horizon remains unresolved \cite{asukula2311,golovnev412024}.

The study of axially symmetric vacuum solutions in NGR, viewed as a generalization of the spherically symmetric case, has been relatively limited. Using the Lorentz-restricted approach, \cite{fukui1981} presented a 1P solution that includes the Kerr solution. The singularities of this Kerr-like spacetime were further analyzed in \cite{toma1991,hecht1992}. Additionally, \cite{bahamonde2021general} examined spherically symmetric vacuum solutions in a generalization of NGR, known as the \( f(\mathscr{V},\mathscr{A},\mathscr{T}) \) theory, where the gravitational action incorporates an arbitrary function \( f \) of the irreducible components of the torsion. Despite these efforts, no non-trivial exact black hole solution has been found, except in cases involving topologically flat horizons or complex-valued tetrads, which are physically problematic \cite{golovnev412024}.

In a proper tetrad (i.e., a gauge with vanishing spin connection), an intuitive explanation has been proposed for why symmetric black holes cannot exist in metric teleparallel theories of gravity, based on the causal structure determined by the metric independent of the specific form of the theory \cite{awad1,golovnev412024}. In such a teleparallel framework, the proper tetrad consists of four covariantly constant vectors, and regardless of the coordinate system, there exists a time-like Killing vector in the exterior region. Parallel transport preserves the covariant constancy of this unit-norm time-like vector field. However, as one approaches the horizon, the direction of this vector tends toward a light-like one, leading to a singularity \cite{golovnev412024}. 

In this study, we investigate static, spherically symmetric vacuum teleparallel geometries within NGR in greater detail. We show that if such a geometry exists in the presence of a local horizon, then either the torsion scalars ($\mathscr{V}$ and $\mathscr{T}$) necessarily diverge there, or the model exhibits some unphysical features.

\subsection{Black holes}

In GR, the geometry is determined by the curvature of the Levi-Civita connection, which is derived from the metric. The spherically symmetric vacuum Schwarzschild solution, governed by Birkhoff’s theorem, is the unique static, spherically symmetric, and asymptotically flat vacuum solution to the Einstein FE, resulting in black holes with horizons that conceal a spacetime singularity.
These types of solutions give rise to intriguing phenomena within the framework of TG, challenging the conventional notions of black holes and gravitational singularities. Unlike in GR, where singularities typically arise only at the origin of the radial coordinate (in Schwarzschild coordinates), TG exhibits divergent behavior near both the horizon and the origin (in Schwarzschild-like coordinates).

In the Schwarzschild manifold, the surface \( r_{s} = 2M \) acts as a horizon, known globally as the event horizon, and is often locally characterized as an apparent horizon (AH) \cite{refAH}. An AH is a marginally outer trapped surface defined by the vanishing expansion of outgoing null geodesics ($\theta_{\sl \ell \sr}=0$), while the ingoing null geodesics remain converging ($\theta_{\sl n \sr} < 0$). Additionally, the expansion of the outgoing null geodesics must decrease when moving inward along the ingoing null direction ($\Delta \theta_{\sl \ell \sr} <0$). These conditions can be summarized as follows~\citep{GH}:
\begin{equation}
\theta_{\sl \ell \sr} = 0 \, , \quad \theta_{\sl n \sr} < 0 \, , \quad \Delta \theta_{\sl \ell \sr} <0 \, .
\end{equation}

This geometric definition underlies the commonly used phenomenological equation for locating the AH, which consists of setting a specific metric function to zero \cite{Hochberg:1998ha}, such as  $g_{tt}=0$. However, this procedure is a coordinate-dependent simplification employed in particular coordinate systems---especially in static, spherically symmetric spacetimes---and it ultimately follows from the ``coordinate-independent" condition \( \theta_{\sl \ell \sr} = 0 \). Here, we do not adopt an invariant definition of a horizon, but instead focus on a local characterization~\citep{GH}. We note, therefore, that \( \theta_{\sl \ell \sr} = 0 \) is a necessary condition for the definition of an AH; such a condition defines the regions we will refer to as a local horizon, as employed in this work.

In TG theories, such as NGR, matter fields are minimally coupled to the metric, as in GR. Consequently, the equations of motion are identical in both frameworks, that is, the test particles follow geodesics. The key difference lies in the geometric interpretation rather than in the physical predictions \citep{aldrovandi1732013}. This allows the definition of a local horizon to be shared between both theories.


%
%
\section{Fundamentals of teleparallel geometry}

The teleparallel geometry is characterized by a tetrad field, \( h^{a}{}_{\mu} \), and a metric-compatible spin connection, \( \omega_{ab\mu} \), which is antisymmetric in the Latin indices (\( \omega_{\sl ab \sr \mu}=0 \)). This framework generates non-zero torsion while maintaining a vanishing curvature \cite{aldrovandi1732013}. The tetrads satisfy the orthogonality conditions \( h^{a}{}_{\mu}h_{a}{}^{\nu} = \delta^{\nu}_{\mu} \) and \( h^{a}{}_{\sigma}h_{b}{}^{\sigma} = \delta^{a}_{b} \), and establish a relationship between the spacetime metric \( g_{\mu\nu} \) and the tangent-space Minkowski metric \( \eta_{ab} = \text{diag}(-1,1,1,1) \), as given by
\begin{equation}
g_{\mu\nu} = h^{a}{}_{\mu}h^{b}{}_{\nu} \eta_{ab} \quad \text{or} \quad \eta_{ab} = h_{a}{}^{\mu}h_{b}{}^{\nu} g_{\mu\nu} \, .
\end{equation}
The total covariant derivative, which acts on both the spacetime and tangent-space indices, vanishes when applied to the tetrad. This condition, known as the \textit{tetrad postulate}~\cite{bahamonde862023}, establishes the relationship between the teleparallel connection, \( \Omega^{\rho}{}_{\nu\mu} \), and spin connection, \( \omega^{a}{}_{b\mu} \), given by
\begin{equation}\label{Omega}
\Omega^{\rho}{}_{\nu\mu} = h_{a}{}^{\rho} \partial_{\mu} h^{a}{}_{\nu} + h_{a}{}^{\rho} \omega^{a}{}_{b\mu} h^{b}{}_{\nu} \, .
\end{equation}
The teleparallel connection features a torsion tensor, defined as
\begin{equation}\label{torsion}
T^{\sigma}{}_{\mu\nu} = 2\Omega^{\sigma}{}_{\al\nu\mu\ar} \, ,
\end{equation}
which is antisymmetric for the last two indices. The torsion tensor can be decomposed into components that are irreducible under the Lorentz group~\cite{mccrea}, commonly referred to as the vector, axial, and purely tensorial parts:
\begin{equation}\label{IPT}
\mathscr{V}_{\mu} = T^{\nu}{}_{\nu\mu} \,,\quad 
\mathscr{A}_{\mu} = \frac{1}{6} \varepsilon_{\mu\nu\rho\sigma} T^{\nu\rho\sigma} \,, \quad 
\mathscr{T}_{\sigma\mu\nu} = T_{\sl\sigma\mu\sr\nu} + \frac{1}{3} \left(g_{\sigma\al\nu} \mathscr{V}_{\mu\ar} + g_{\mu\al\nu} \mathscr{V}_{\sigma\ar} \right) \, .
\end{equation}
Here, the axial torsion is defined using the totally antisymmetric Levi-Civita tensor \( \varepsilon_{\mu\nu\rho\sigma} \) associated with the metric \( g_{\mu\nu} \).

A fundamental result stemming from the metric compatibility of Lorentz connections is that any such connection can be decomposed into the Levi-Civita connection $ \Gamma^{\sigma}{}_{\mu\nu}$ and the contortion tensor \cite{aldrovandi1995introduction}:
\begin{equation}\label{ConDec}
\Omega^{\sigma}{}_{\mu\nu} = \Gamma^{\sigma}{}_{\mu\nu} + K^{\sigma}{}_{\mu\nu} \, ,
\end{equation}
where the contortion tensor, which is antisymmetric in the first two indices, is given by
\begin{equation}\label{Cont}
K_{\sigma\mu\nu} = T_{\al\mu\sigma\ar\nu}+\frac{1}{2} T_{\nu\sigma\mu} \, .
\end{equation}
An important consequence of decomposition \eqref{ConDec} is that the curvature tensor associated with the Levi-Civita connection can be expressed as
\begin{equation}\label{RimanToN}
\mathcal{R}^{\rho}{}_{\sigma\mu\nu} = 2N^{\rho}{}_{\sigma\al\mu\nu\ar} \, ,
\end{equation}
where
\begin{equation}
N^{\rho}{}_{\sigma\mu\nu} \equiv \nabla_{\nu} K^{\rho}{}_{\sigma\mu} + K^{\rho}{}_{\lambda\mu} K^{\lambda}{}_{\sigma\nu} + K^{\rho}{}_{\sigma\lambda} K^{\lambda}{}_{\mu\nu} \, ,
\end{equation}
which is a tensor that features antisymmetry in its first two indices and \(\nabla\) denotes the covariant derivative with respect to the teleparallel connection \eqref{Omega}.
Contracting indices in Eq.~\eqref{RimanToN} yields the Ricci scalar, which satisfies the following identity:
\begin{equation}
\mathcal{R} = 2N^{\mu\nu}{}_{\al\mu\nu\ar} = -T - \frac{2}{h} \partial_{\nu} (h T^{\mu\nu}{}_{\mu}) \, ,
\end{equation}
with
\begin{equation}
T = K^{\mu\nu\rho} K_{\rho\nu\mu} - K^{\mu\rho}{}_{\mu} K^{\nu}{}_{\rho\nu} \, .
\end{equation}
This scalar, known as the torsion scalar, plays a central role in the Lagrangian formulation of teleparallel theories. It can also be obtained from
\begin{equation}\label{TdeS}
T = \frac{1}{2} S_{\rho}{}^{\mu\nu} T^{\rho}{}_{\mu\nu} \, ,
\end{equation}
where \( S_{\rho}{}^{\mu\nu} \), the superpotential, is antisymmetric in the last two indices and takes the form
\begin{equation}\label{S}
S_{\rho}{}^{\mu\nu} = K^{\mu\nu}{}_{\rho} + \delta^{\mu}_{\rho} T^{\lambda\nu}{}_{\lambda} - \delta^{\nu}_{\rho} T^{\lambda\mu}{}_{\lambda} \, .
\end{equation}
An alternative expression for the invariant in \eqref{TdeS} can be written in terms of the irreducible components of the torsion tensor in \eqref{IPT} as
\begin{equation}\label{TorS2}
T = \frac{3}{2} \mathscr{A} - \frac{2}{3} \mathscr{V} + \frac{2}{3} \mathscr{T} \, ,
\end{equation}
where
\begin{equation}\label{Tds}
\mathscr{A} = \mathscr{A}^{\mu} \mathscr{A}_{\mu} \,, \quad 
\mathscr{V} = \mathscr{V}^{\mu} \mathscr{V}_{\mu} \,, \quad 
\mathscr{T} = \mathscr{T}^{\sigma\mu\nu} \mathscr{T}_{\sigma\mu\nu} \, .
\end{equation}

To investigate static and spherically symmetric solutions in teleparallel geometry, one must first specify the general forms of the frame (or co-frame) and the flat, metric-compatible spin connection that is compatible with the underlying static and spherically symmetric affine symmetries \cite{coley616439,coley842024}.
\subsection{Static and spherically symmetric teleparallel geometries}

Working in coordinates $x^{\mu}=(t,r,\theta,\phi)$ and using the affine frame symmetry generators of the three-dimensional spherical symmetry group and the additional affine static spherically symmetric generator $\partial_{t}$ \cite{coley616439}, the most general class of these geometries was derived by ``diagonalizing" the frame into a tetrad defined by three arbitrary functions: $A_{1}(r)$, $A_{2}(r)$, and $A_{3}(r)$. That is,
\begin{equation}\label{TetradSS}
h^{a}{}_{\mu}=
\begin{pmatrix}
A_1 & 0 & 0 &0\\
0 & A_2 & 0 & 0 \\
0 & 0 & A_3 & 0 \\
0 & 0 & 0 & A_3\,\sin \theta
\end{pmatrix} \, .
\end{equation}
This invariant symmetry frame allows the derivation of the most general metric-compatible connection. By imposing the flatness condition, any static spherically symmetric teleparallel geometry can be described using the three functions in (\ref{TetradSS}), along with two additional functions, $\chi(r)$ and $\psi(r)$, in the spin connection \cite{coley616439}, whose non-zero components are given by:
\begin{alignat}{1}\label{SpinV}\nonumber
&\omega_{133}=\omega_{144}= \cos\chi \sinh\psi /A_{3} \, , \quad  \omega_{134}=\omega_{413}=\sin\chi \sinh\psi /A_{3}  \, ,\\[1.5ex]
& \omega_{234}=\omega_{423}=\sin\chi \cosh\psi/A_{3} \, , \quad  \omega_{233}=\omega_{244} = \cos\chi\cosh\psi/A_{3}\,,\\[1.5ex]\nonumber
&\omega_{212}= \psi'/A_{2} \, , \quad \omega_{432}= \chi'/A_{2} \, , \quad \omega_{434}= \cot\theta/A_{3} \, .
\end{alignat}
It is important to note that we still have coordinate freedom to choose the radial component of the frame whenever $A_{3}$ is not constant. Additionally, any choice of the arbitrary functions $\chi$ and $\psi$ uniquely determines the teleparallel geometry, as any alteration in the spin connection that affects the form of $\chi$ or $\psi$ also changes the tetrad \cite{coley616439}.

In \cite{robert2024} it was shown that when the tetrad (\ref{TetradSS}) and spin connection (\ref{SpinV}) undergo a Lorentz transformation given by
\begin{equation}
\Lambda^{a}{}_{b}=\begin{pmatrix}
\cosh\psi & \sinh\psi\cos\phi\sin\theta & \sinh\psi\sin\theta\sin\phi & \sinh\psi\cos\theta \\
\sinh\psi & \cosh\psi\cos\phi\sin\theta & \cosh\psi\sin\theta\sin\phi & \cosh\psi\cos\theta \\
0 & \sin\chi\sin\phi-\cos\chi\cos\theta\cos\phi &-\sin\chi\cos\phi -\cos\chi\cos\theta\sin\phi & \cos\chi\sin\theta \\
0 & \cos\chi\sin\phi+\sin\chi\cos\theta\cos\phi & -\cos\chi\cos\phi+\sin\chi\cos\theta\sin\phi & -\sin\chi\sin\theta
\end{pmatrix} \, ,
\end{equation}
the pair is transformed into the proper frame, where ${\omega'}^{a}{}_{b\mu}=0$ and ${h'}^{a}{}_{\mu}=\Lambda^{a}{}_{b}h^{b}{}_{\mu}$. We write the explicit form of the tetrad in columns because of the length of the expression as follows:
\begin{equation}\label{hpss}
\begin{alignedat}{2} 
{h'}^{1}{}_{t} & = A_{1}\cosh\psi \,, && \quad \quad h'^{1}{}_{\theta}  = 0\,, \\
h'^{2}{}_{t} & = A_{1}\sinh\psi\sin\theta\cos\phi \,,&& \quad \quad h'^{2}{}_{\theta}  = A_{3}(\sin\chi\sin\phi-\cos\chi\cos\theta\cos\phi)\,, \\
h'^{3}{}_{t} & = A_{1}\sinh\psi\sin\theta\sin\phi \,, && \quad \quad h'^{3}{}_{\theta}  = -A_{3}(\sin\chi\cos\phi+\cos\chi\cos\theta\sin\phi)\,, \\
h'^{4}{}_{t} & = A_{1}\sinh\psi\cos\theta \,,&& \quad \quad h'^{4}{}_{\theta}  = A_{3}\cos\chi\sin\theta\,, \\
h'^{1}{}_{r} & = A_{2}\sinh\psi\,, && \quad \quad h'^{1}{}_{\phi}  = 0\,, \\
h'^{2}{}_{r} & =A_{2}\cosh\psi\sin\theta\cos\phi\,, && \quad \quad h'^{2}{}_{\phi}  =A_{3}\sin\theta(\sin\chi\cos\theta\cos\phi+\cos\chi\sin\phi)\,, \\
h'^{3}{}_{r} & =A_{2}\cosh\psi\sin\theta\sin\phi\,, && \quad \quad h'^{3}{}_{\phi}  =A_{3}\sin\theta(\sin\chi\cos\theta\sin\phi-\cos\chi\cos\phi) \,,\\
h'^{4}{}_{r} & =A_{2}\cosh\psi\cos\theta \,,&& \quad \quad h'^{4}{}_{\phi}  =-A_{3}\sin\chi\sin^{2}\theta \, .
\end{alignedat}
\end{equation}
Note that both the diagonal and proper frames satisfying spherical symmetry have four degrees of freedom: two \(\{A_{1}, A_{2}\}\) are associated with frame symmetry, and two \(\{\chi, \psi\}\) correspond to Lorentz rotations and boosts, respectively \cite{robert2024}. This is because of the remaining freedom in choosing coordinates and redefining \(A_{3}\).
\subsection{Teleparallel scalars}

The teleparallel invariants introduced in Eq.~\eqref{Tds} play a crucial role in the analysis of static and spherically symmetric teleparallel geometries. Because of their presence in the Lagrangian formulations of teleparallel theories, these invariants can help determine the nature of special surfaces characterized by the explicit forms of the arbitrary functions \( A_{1} \), \( A_{2} \), \( \chi \), and \( \psi \). By substituting the tetrad \eqref{TetradSS} and spin connection \eqref{SpinV}, the teleparallel scalars were found to be
\begin{subequations}\label{TScalars}
\begin{alignat}{1}
\mathscr{V} & =
4\left(\frac{\cos\chi}{A_{3}}\right)^2+\frac{4\cos\chi}{A_{2}A_{3}}\left([\ln(A_{1}A_{3}{}^{2})]'\cosh\psi+[\cosh\psi]' \right) +\left(\frac{[\ln (A_{1}A_{3}{}^{2})]'}{A_{2}}\right)^{2}-\left(\frac{\psi'}{A_{2}} \right)^{2} , \\[11pt]
\mathscr{A} & =
-\left(\frac{4 \sin \chi}{3 A_3}\right)^2
- \frac{16 \cosh \psi [\cos \chi]'}{9 A_2 A_3}
- \left(\frac{2\chi'}{3A_{2}} \right)^{2} , \\[11pt]
\mathscr{T} & =
\frac{1}{A_3^2}+\frac{[\ln(A_{1}/A_{3})]'}{A_{2}}\left(\frac{[\ln(A_{1}/A_{3})]'}{A_{2}}-\frac{2\cos\chi\cosh\psi}{A_{3}}\right) -\frac{2[\cos\chi\cosh\psi]'}{A_{2}A_{3}}+\frac{(\chi')^2-(\psi')^2}{A_{2}{}^{2}}.
\end{alignat}
\end{subequations}
It is also useful to consider the linear combination of these scalars that yields the torsion scalar \eqref{TorS2} in terms of the arbitrary functions defining the static and spherically symmetric teleparallel geometry:
\begin{alignat}{1} \label{TS1}
T & = -\frac{2}{A_{2}{}^{2}}\left(\frac{A_{2}{}^{2}}{A_{3}{}^{2}}+[\ln A_{3}]'[\ln (A_{1}{}^{2}A_{3})]'\right)-\frac{4}{A_{2}A_{3}}\left([\ln (A_{1}A_{3})]'\cos\chi\cosh\psi+[\cos\chi\cosh\psi]'\right) \, .
\end{alignat}

To describe a well-defined black-hole geometry, the invariants entering any teleparallel Lagrangian must remain finite at the local horizon. If any of these invariants diverge at that location, this implies that the local horizon and its interior are not part of the manifold. This contradicts the definition of a black hole, where the only region excluded from the manifold is the true physical singularity.
%
%
\section{Expansion of null congruences}\label{ENC}
In the teleparallel framework, particle dynamics are governed by force-like equations that correspond to the geodesics of GR~\citep{aldrovandi1732013}. These equations take the following forms:  
\begin{equation}\label{geo}
u^{\nu} \nabla_{\nu} u^{\mu} = K^{\mu}{}_{\sigma \nu} u^{\sigma} u^{\nu} \, ,
\end{equation}  
where \( u^{\mu} \) denotes the four-velocity of the particle. Let us now consider a congruence of null geodesics, whose expansion describes how a bundle of nearby null rays diverges or converges as it evolves along the geodesic direction. Let \( \ell^{\mu} \) and \( n^{\mu} \) be outgoing and ingoing null tangent vector fields to the geodesics, respectively. These vectors satisfy the null condition and are mutually normalized as follows:
\begin{equation}\label{lncond}
\ell^{\mu}\ell_{\mu} = n^{\mu}n_{\mu} = 0 \quad \text{and} \quad \ell^{\mu}n_{\mu} = -1\, .
\end{equation}
Note that these vectors also satisfy Eq.~\eqref{geo}, as well as the identities
\begin{equation}\label{nbln}
\ell^{\nu} \nabla_{\mu} \ell_{\nu} = n^{\nu} \nabla_{\mu} n_{\nu} = 0 \, .
\end{equation}
The induced transverse metric, orthogonal to both \( \ell^{\mu} \) and \( n^{\mu} \), is given by  
\begin{equation}\label{trmetric}
m_{\mu\nu} = g_{\mu\nu} + 2 \ell_{\sl\mu} n_{\nu\sr} \, .
\end{equation}
Since the normalization conditions in Eq.~\eqref{lncond} does not uniquely determine \( n^{\mu} \) given \( \ell^{\mu} \), or vice versa; consequently, the transverse metric is  not unique \citep{poisson}. Using the null vectors and transverse metric, we define the expansion scalars associated with the outgoing and ingoing directions as follows:
\begin{equation}\label{expansion}
\theta_{\sl \ell \sr} = m^{\mu\nu} \left( \nabla_{\mu} \ell_{\nu} + K^{\sigma}{}_{\nu\mu} \ell_{\sigma} \right) 
\quad \text{and} \quad 
\theta_{\sl n \sr} = m^{\mu\nu} \left( \nabla_{\mu} n_{\nu} + K^{\sigma}{}_{\nu\mu} n_{\sigma} \right)\, ,
\end{equation}
where \( \theta_{\sl \ell \sr} \) measures the expansion of outgoing null geodesics, and \( \theta_{\sl n \sr} \) measures the expansion of the ingoing null geodesics. The interpretation of these scalars is as follows: when positive, the associated geodesics diverge; when negative, they converge; and when equal to zero, they indicate stationary behaviour (marginal expansion)\citep{hawking}. An alternative form of the expressions in Eq.~\eqref{expansion} follows from Eqs.~\eqref{geo}, \eqref{nbln}, and \eqref{trmetric}, leading to  
\begin{equation}\label{expansionf}
\theta_{\sl \ell \sr} = \nabla_{\mu} \ell^{\mu} + K^{\sigma\mu}{}_{\mu} \ell_{\sigma} 
\quad \text{and} \quad 
\theta_{\sl n \sr} = \nabla_{\mu} n^{\mu} + K^{\sigma\mu}{}_{\mu} n_{\sigma} \, .
\end{equation}
These expressions show that, although \( m^{\mu\nu} \) depends on both the outgoing and ingoing null vectors, each expansion scalar depends only on its associated null vector.

An important application of Eq.~\eqref{expansionf} is the definition of an AH, which is a marginally outer trapped surface—specifically, the outermost 2-surface satisfying  
\begin{equation}\label{AHD}
\theta_{\sl \ell \sr} = 0 \, , \quad \theta_{\sl n \sr} < 0 \, , \quad \Delta \theta_{\sl \ell \sr}= n^{\mu}\nabla_{\mu}\theta_{\sl \ell \sr} < 0 \, .
\end{equation}
Note that the condition \( \theta_{\sl \ell \sr} = 0 \) must hold precisely on the AH, thereby identifying it as a marginally outer trapped surface. The condition \( \theta_{\sl n \sr} < 0 \) also holds on the AH; although it does not define the AH, it ensures the surface bounds a trapped region rather than a degenerate or marginally inner-trapped region. Finally, the expansion \( \theta_{\sl \ell \sr} \) must decrease along the ingoing null direction, \( \Delta \theta_{\sl \ell \sr} < 0 \), in a neighbourhood immediately inside the AH. This guarantees that the AH locally encloses a region where outgoing null rays are already converging, thereby providing a stability criterion~\citep{hawking}.

In static and spherically symmetric teleparallel geometries determined by Eqs.~\eqref{TetradSS} and \eqref{SpinV}, the outgoing and ingoing null vectors can be obtained by using the conditions in Eq.~\eqref{lncond}, and are expressed in terms of the arbitrary functions \( A_{1} \) and \( A_{2} \) as follows~\citep{coley616439}:
\begin{equation}\label{null}
\boldsymbol{\ell} = \frac{1}{\sqrt{2}}\left( \frac{1}{A_{1}}, \frac{1}{A_{2}}, 0, 0 \right) 
\quad \text{and} \quad 
\boldsymbol{n} = \frac{1}{\sqrt{2}}\left( \frac{1}{A_{1}}, -\frac{1}{A_{2}}, 0, 0 \right) \, .
\end{equation}
This choice of null vectors is symmetric under time reversal and radial reflection; that is, they share the same temporal component, whereas their radial components mirror each other in the radial direction. By substituting the expressions in \eqref{null} into Eq.~\eqref{expansionf}, the expansion scalars were found to be
\begin{equation}\label{theAH}
\theta_{\sl \ell \sr}=-\theta_{\sl n \sr} = \frac{\sqrt{2}}{A_{2}}[\ln A_{3}]' \, .
\end{equation}
Note that these scalars are equal in magnitude and opposite in sign, as expected for static and spherically symmetric spacetimes when a symmetric choice of null vectors such as those in \eqref{null} is used. This implies that the expansion scalars share the same roots, that is, $\theta_{\sl \ell \sr}(r_{h}) = \theta_{\sl n \sr}(r_{h}) = 0$, where $r = r_{h}$ denotes the location of a possible AH. Consequently, the condition $\theta_{\sl n \sr} < 0$ in \eqref{AHD} is only valid outside the region $r = r_{h}$ in this specific case. Then, a more appropriate formulation of the AH definition in the context of static and spherically symmetric geometries, when symmetric null vectors are used, is then given by
\begin{equation}\label{AHD2}
\theta_{\sl \ell \sr} = -\theta_{\sl n \sr}= 0 \, , \quad \Delta \theta_{\sl \ell \sr}< 0 \, .
\end{equation}
This highlights the role of the directional derivative of the outgoing expansion scalar along the ingoing null direction in characterizing the AH. By computing this explicitly, we obtain
\begin{equation}\label{DthD}
\Delta \theta_{\sl \ell \sr} = \frac{1}{A_{2}^{2}}\left([\ln A_{2}]'[\ln A_{3}]' - [\ln A_{3}]'' \right) \, .
\end{equation}
Both \eqref{AHD2} and \eqref{DthD} depend solely on \( A_{2} \), because \( A_{3} \) remains freely specified through an appropriate choice of the coordinate gauge. In this study, we only used the first condition in \eqref{AHD2}, namely \( \theta_{\sl \ell \sr} = 0 \). This provides a necessary condition for the existence of an AH, which we refer to as the local horizon.

%
%
\section{New general relativity}
The class of theories known as NGR constitutes the first and simplest modification of TEGR explored in the literature \cite{hayashi191981}. In NGR, the coefficients associated with the irreducible torsion scalars are treated as three independent free parameters, \( (c_{v}, c_{a}, c_{t}) \), which are expected to be constrained by the experimental data. These parameters allow for deviations from the TEGR Lagrangian, corresponding to the specific values \( (c_{v} = -2/3, c_{a} = 3/2, c_{t} = 2/3) \), thereby enabling NGR to incorporate corrections to the standard gravitational predictions provided by TEGR \cite{bahamonde862023}. The general NGR Lagrangian density can be written as:
\begin{equation}\label{Lngr}
\mathcal{L} = c_a \mathscr{A} + c_t \mathscr{T} + c_v \mathscr{V} \, ,
\end{equation}
and the corresponding NGR action, including a matter Lagrangian density \( \mathcal{L}_{m} \), is given by
\begin{equation}\label{S}
\mathcal{S} = \int h \left( \kappa\mathcal{L} + \mathcal{L}_{m} \right) d^{4}x \, ,
\end{equation}
where \( h \) denotes the determinant of the tetrad and \( \kappa \) is a gravitational coupling constant. Note that the action \eqref{S} is well-defined only in regions of the manifold where the torsion scalars \( \mathscr{V} \), \( \mathscr{A} \), and \( \mathscr{T} \) appearing in the Lagrangian \eqref{Lngr} remain finite.

Variation of the action with respect to the tetrad yields the FE:
\begin{alignat}{1} \label{FE}\nonumber
W_{\mu\nu} & \equiv c_{v}\left( 2g_{\mu\al\nu}\nabla^{\rho}\mathscr{V}_{\rho\ar} - \frac{1}{2}g_{\mu\nu}\mathscr{V} \right) + c_{a}\left( \frac{1}{6}g_{\mu\nu}\mathscr{A} + \frac{1}{3}\mathscr{A}_{\mu}\mathscr{A}_{\nu} \right. \\[0.5ex]\nonumber
\quad & \left. + \frac{2}{9}\mathscr{A}^{\alpha} \left( 2\epsilon_{\alpha\nu\rho\sigma}\mathscr{T}_{\mu}{}^{\rho\sigma} - \epsilon_{\mu\nu\rho\sigma}\mathscr{T}_{\alpha}{}^{\rho\sigma} \right) + \frac{1}{3}\epsilon_{\mu\nu\rho\sigma} (\mathscr{V}^{\rho}\mathscr{A}^{\sigma} - \nabla^{\rho}\mathscr{A}^{\sigma}) \right) + c_{t}\left( \frac{1}{3}g_{\mu\nu}\mathscr{T} \right. \\[0.5ex]
\quad & \left. - \frac{1}{3}g_{\mu\nu} \mathscr{T}_{\alpha\sigma\rho} \mathscr{T}^{\alpha\rho\sigma} + \frac{8}{3} \mathscr{T}^{\rho}{}_{\al\sigma\nu\ar} \mathscr{T}^{\sigma}{}_{\al\mu\rho\ar} - \mathscr{V}^{\rho}\mathscr{T}_{\mu\al\nu\rho\ar} + 2\nabla^{\rho} \mathscr{T}_{\mu\al\nu\rho\ar} + 2 \mathscr{A}^{\alpha} \epsilon_{\alpha\sigma\rho\sl\mu} \mathscr{T}_{\nu\sr}{}^{\rho\sigma} \right) = \kappa \Theta_{\mu\nu} \, ,
\end{alignat}
where the energy-momentum tensor is defined as
\begin{equation}\label{momento}
\Theta_{\mu\nu} = -h^{-1} \frac{\delta(h \mathcal{L}_{m})}{\delta h^{a}{}_{\rho}} h^{a}{}_{\mu} \, g_{\rho\nu} \, .
\end{equation}

The variation of the action~\eqref{S} with respect to a non-symmetric tetrad generally leads to non-symmetric FE~\eqref{FE}. When variation is performed with respect to the spin connection, it yields FE equivalent to the antisymmetric part of the tetrad FE~\cite{bahamonde862023}. This reflects the underlying local Lorentz invariance of the theory, with the antisymmetric part imposing consistency conditions that ensure covariance. If the source~\eqref{momento} is symmetric, for instance, it follows that the antisymmetric part of the FE~\eqref{FE} must vanish as follows
\begin{equation}\label{Wa}
W_{\al\mu\nu\ar} = 0 \, .
\end{equation}
In the TEGR case, where \( c_{v} = -2/3 \), \( c_{a} = 3/2 \), and \( c_{t} = 2/3 \), Eq.~\eqref{Wa} is satisfied automatically. However, in NGR, fulfilling condition \eqref{Wa} becomes a non-trivial task.

In the followings, we present the antisymmetric and symmetric components of the vacuum NGR FE using the ansatz defined in Eqs.~\eqref{TetradSS} and \eqref{SpinV}. To simplify the coefficient expressions and facilitate the algebraic treatment of the FE, we introduced a new set of parameters:
\begin{equation}\label{bp}
b_1 = c_t + c_v, \quad b_2 = 3c_t, \quad b_3 = -\frac{4c_a}{9} - c_v \, .
\end{equation}
Note that the parameters \( b_{1} \) and \( b_{3} \) are coupled through \( c_{v} \), whereas \( b_{1} \) and \( b_{2} \) are coupled through \( c_{t} \).

\subsection{Antisymmetric FE}\label{AFE}

The antisymmetric part of the FE (AFE) \eqref{FE} is given by
\begin{alignat}{1}\label{FEA}\nonumber
W_{\al\mu\nu\ar} & = c_{v} \nabla^{\rho} \mathscr{V}_{[\mu} g_{\nu]\rho}
- \frac{c_{a}}{3} \left( \frac{2}{3} \left( \epsilon_{\mu\nu\rho\gamma} \mathscr{T}_{\sigma}{}^{\rho\gamma} - 2\epsilon_{\sigma\rho\gamma[\nu} \mathscr{T}_{\mu]}{}^{\rho\gamma} \right)\mathscr{A}^{\sigma} 
+ \epsilon_{\mu\nu\sigma\rho} (\mathscr{V}^{\rho}\mathscr{A}^{\sigma}  - \nabla^{\rho}\mathscr{A}^{\sigma} ) \right)  \\[0.5ex]
\quad & + \frac{c_{t}}{2} (\mathscr{V}^{\rho}\mathscr{T}_{\rho\al\mu\nu\ar} - 2\nabla^{\rho}\mathscr{T}_{\rho\al\mu\nu\ar})  = 0 \, .
\end{alignat}
These equations are not automatically satisfied because of the presence of the three free parameters of the theory ($c_{v},c_{t},c_{a}$), which generally prevent the terms from cancelling out, as they do in special cases such as
\begin{equation}
b_{1} = b_{3} = 0 \, ,
\end{equation}
which corresponds to a constant multiple (or rescaled version) of TEGR \cite{bahamonde862023}. Using the reparameterization introduced in Eq.~\eqref{bp} and the static and spherically symmetric ansatz \eqref{hpss}, the AFE \eqref{FEA} reduce to
\begin{subequations}\label{FEAV}
\begin{alignat}{1} \label{FEAVa}
W_{ \al 12 \ar} & : \, \frac{b_{1}}{2} \left[(A_{1}A_{3}{}^{2}/A_{2})\psi'\right]' 
+ A_{1}A_{3}\left(b_{1}[\ln A_{3}]'\cos\chi + b_{3}[\cos\chi]' \right) \sinh\psi = 0 \, , \\[2 ex]
W_{ \al 34 \ar} & : \, \frac{1}{2}(b_{1}+b_{3}) \left[(A_{1}A_{3}{}^{2}/A_{2})\chi'\right]'  
- A_{1}A_{3}\left(F_{0}\cosh\psi + b_{3} \, [\cosh\psi]' + 2b_{3}\frac{A_{2}}{A_{3}}\cos\chi \right)\sin\chi = 0 \, ,
\end{alignat}
\end{subequations}
where
\begin{equation}
F_{0}= b_{3}\,[\ln A_{1}]'-(b_{1}-b_{3})[\ln A_{3}]'\, .
\end{equation}
Notably, Eqs.~\eqref{FEAV} do not involve the parameter \( b_{2} \) and cannot be readily solved for \( \chi \) and \( \psi \), mainly because \( A_{1} \) and \( A_{2} \) remain unspecified. A useful strategy for analyzing potential solutions of Eqs.~\eqref{FEAV} is to assume that \( \chi \) and \( \psi \) are constants, denoted \( \chi = \chi_{\n} \) and \( \psi = \psi_{\n} \). Under this assumption, the AFE are reduced to
\begin{subequations}\label{AFEcp}
\begin{alignat}{1}
W_{ \al 12 \ar} & : \, A_{1}A_{3}\left(b_{1}[\ln A_{3}]'\cos\chi_{\n} \right)  \sinh\psi_{\n}=0 \, , \\[2 ex]
W_{ \al 34 \ar} & : \, A_{1}A_{3}\left(F_{0}\cosh\psi_{\n} + 2b_{3}\frac{A_{2}}{A_{3}}\cos\chi_{\n} \right)\sin\chi_{\n}=0 \, ,
\end{alignat}
\end{subequations}
Evidently, the choices \( \psi_{\n} = 0 \) and \( \chi_{\n} = n\pi \) render both Eqs.~\eqref{FEAV} identically zero. However, this is not the only possible solution. Since the parameters \( b_{1} \) and \( b_{3} \) remain free, alternative scenarios may also satisfy the AFE. For instance, when \( b_{1} = 0 \) and \( \chi_{\n} = n\pi \), Eqs.~\eqref{FEAV} are satisfied for any arbitrary function \( \psi(r) \). This naturally raises the question of whether non-constant \( \chi \) and \( \psi \) can also satisfy the AFE under minimal restrictions on the parameters \( b_{1} \) and \( b_{3} \), and what constraints this might impose on the arbitrary functions \( A_{1} \) and \( A_{2} \); we explore this further in Section~\ref{pertuA}. 

Furthermore, in the TEGR limit, where \( b_{1} = b_{3} = 0 \), the Eqs.~\eqref{FEAV} hold for arbitrary functions \( \chi(r) \) and \( \psi(r) \), implying a certain degeneracy. This is an issue which we shall return in future work.

\subsection{Symmetric FE in vacuum}

The symmetric part of the FE (SFE) \eqref{FE} in vacuum is given by
\begin{alignat}{1}\label{FES}\nonumber
W_{\sl\mu\nu\sr}& =c_{v}\left(-\frac{1}{2}g_{\mu\nu}\mathscr{V} +g_{\mu\nu}\nabla^{\rho}\mathscr{V}_{\rho}-\nabla^{\rho}\mathscr{V}_{\sl\mu}g_{\nu\sr\rho} \right)+c_{a}\left(\frac{1}{6}g_{\mu\nu}\mathscr{A}+\frac{1}{3}\mathscr{A}_{\mu}\mathscr{A}_{\nu} \right. \\[0.5ex]\nonumber
\quad & \left. +\frac{4}{9}\epsilon_{\sigma\rho\gamma\sl\mu}\mathscr{T}_{\nu\sr}{}^{\rho\gamma}\mathscr{A}^{\sigma}\right)+c_{t}\left(\frac{1}{3}g_{\mu\nu}(\mathscr{T}-\mathscr{T}_{\alpha\sigma\rho}\mathscr{T}^{\alpha\rho\sigma})+\frac{8}{3}\mathscr{T}^{\rho}{}_{\al\sigma\nu\ar}\mathscr{T}^{\sigma}{}_{\al\mu\rho\ar} \right.\\[0.5ex]
\quad & \left. -2\epsilon_{\sigma\rho\alpha\sl\mu}\mathscr{T}_{\nu\sr}{}^{\rho\alpha}\mathscr{A}^{\sigma} +\nabla^{\rho}(\mathscr{T}_{\mu\nu}{}{}^{\sigma}-\mathscr{T}^{\sigma}{}_{\sl\mu\nu\sr})-\frac{1}{2}\mathscr{V}^{\rho}(\mathscr{T}_{\mu\nu}{}{}^{\sigma}-\mathscr{T}^{\sigma}{}_{\sl\mu\nu\sr}) \right) = 0\, .
\end{alignat}
We now focus on its explicit form in the static and spherically symmetric case, for which the SFE exhibit four non-trivial components: three independent diagonal components \((1,1)\), \((2,2)\), and \((3,3)\), along with one off-diagonal component \((1,2)\), given by:
\begin{subequations}\label{FESV}
\begin{alignat}{1}\nonumber
W_{\sl 11 \sr} &: \, F_{1} -\frac{1}{2}(b_{1}+b_{3})(\chi')^{2}+\frac{b_{1}}{2}(\psi')^{2}-2b_{3}\left(\frac{A_{2}}{A_{3}}\sin\chi\right)^{2}\\[1.5ex]
&\quad-2\frac{A_{2}}{A_{3}}\left(  b_{1}[\ln A_{3}]'\cos\chi+b_{3}[\cos\chi]'\right)\cosh\psi=0 \, , \\[2ex] \label{FESV22}
W_{ \sl 22 \sr} &: F_{2}-\frac{1}{2}(b_{1}+b_{3})(\chi')^{2} +\frac{b_{1}}{2}(\psi')^{2} +2 b_{3}\left(\frac{A_{2}}{A_{3}}\sin\chi\right)^{2} =0\, ,\\[2ex]\nonumber
W_{ \sl 33 \sr} &: \, F_{3}+\frac{1}{2}(b_{1}+b_{3})(\chi')^{2}-\frac{b_{1}}{2}(\psi')^{2}-b_{1}\frac{A_{2}}{A_{3}}[\cosh\psi]'\cos\chi \\[1.5ex]
& \quad -\frac{A_{2}}{A_{3}}\left(b_{1}[\ln A_{1}]'\cos\chi+(b_{1}-b_{3})[\cos\chi]' \right)\cosh\psi=0 \, , \\[2ex]\label{FESVd}
W_{ \sl 12 \sr} & : \, -\frac{b_{1}}{2} \left[(A_{1}A_{3}{}^{2}/A_{2})\psi'\right]'- A_{1}A_{3}\left(b_{1}[\ln A_{3}]'\cos\chi +b_{3}[\cos\chi]' \right)  \sinh\psi=0 \, .
\end{alignat}
\end{subequations}
Here, a set of functions \( F_{i} \) is introduced, each depending solely on the arbitrary functions \( A_{j} \); that is, neither \( \chi \) nor \( \psi \) appears in any of the terms. These functions are given by

\begin{subequations}\label{FV}
\begin{alignat}{1}\nonumber
F_{1}& =  b_{1}[\ln A_{1}]''+(2b_{1}-b_{2})[\ln A_{3}]''+b_{1}[\ln A_{1}]'[\ln (A_{1}{}^{1/2}A_{3}{}^{2}/A_{2})]'\\[1.5ex] 
& \quad +[\ln A_{3}]'\left((b_{2}-2b_{1})[\ln A_{2}]' +(2b_{1}-3b_{2}/2)[\ln A_{3}]'\right)-\left(2b_{1}-b_{2}/2\right)\left(\frac{A_{2}}{A_{3}}\right)^{2}  \, , \\[2ex]\nonumber 
F_{2}&= \, -\frac{b_{1}}{2}\left([\ln A_{1}]'\right)^{2} +[\ln A_{3}]'\left((b_{2}-2b_{1})[\ln A_{1}]' +(b_{2}/2-2b_{1})[\ln A_{3}]'\right)
\\[1.5ex]
&\quad +\left(2b_{1}-b_{2}/2 \right)\left(\frac{A_{2}}{A_{3}}\right)^{2}\, ,\\[2ex]\nonumber
F_{3}&= \, (b_{2}/2-b_{1})[\ln A_{1}]'' -(2b_{1}-b_{2}/2)[\ln A_{3}]'' -(2b_{1}-b_{2}/2)[\ln A_{3}]'[\ln (A_{1}A_{3}/A_{2})]'  \\[1.5ex] 
& \quad + \frac{1}{2}[\ln A_{1}]'\left((b_{2}-b_{1})[\ln A_{1}]'+(2b_{1}-b_{2})[\ln A_{2}]' \right)\,  .
\end{alignat}
\end{subequations}
Note that Eq.~\eqref{FESVd} differs from Eq.~\eqref{FEAVa} by an overall negative sign. Consequently, the system \eqref{FESV} contains only three independent SFE, since satisfying Eq.~\eqref{FEAVa} guarantees that Eq.~\eqref{FESVd} is also satisfied. Moreover, fulfilling the condition \eqref{Wa} is a necessary prerequisite before solving the SFE.

The four FE in~\eqref{FESV} involve the four variables \( \{A_{1}, A_{2}, \chi, \psi\} \), since the coordinate freedom can be used to fix \( A_{3} \).  In addition, the AFE impose further constraints on \( \chi \) and \( \psi \), effectively reducing the number of independent variables to two and simplifying the problem of finding solutions. Nonetheless, solving the SFE remains challenging when \( \chi \) and \( \psi \) are treated as general functions of \( r \). Even in the simplified case where \( \chi \) and \( \psi \) are taken as constants, finding explicit solutions depends heavily on the choice of \( A_{3} \). In particular, we show that this task becomes more tractable in isotropic coordinates, for which \( A_{3} = rA_{2} \), as demonstrated in Appendix~\ref{app}, where the exact solution found in~\citep{obukhov672003} is reviewed and analysed.

\subsection{NGR normalization}

Consider the original form of the Lagrangian for NGR \eqref{Lngr}. The issue with the Lagrangian \eqref{Lngr} is that, for certain values of the parameters \( c_a \), \( c_t \), and \( c_v \), it becomes simply a rescaled version of TEGR. Analyzing this Lagrangian reveals two distinct sets of models to consider:
\begin{equation}
c_v = 0 \quad \text{and} \quad c_v \neq 0.
\end{equation}
In the first case where \( c_v = 0 \), TEGR is not recovered in any limit, and the theory reduces to a 2P model given by:
\begin{equation}\label{Lnon}
\mathcal{L} = c_a \mathscr{A} + c_t \mathscr{T}.
\end{equation}
In the second case, \( c_v \neq 0 \). We begin by rewriting the NGR Lagrangian \eqref{Lngr} in the form:
\begin{equation}\label{Lref}
\mathcal{L} = \left( c_a + \frac{9}{4} c_v \right) \mathscr{A} + (c_t + c_v) \mathscr{T} - \frac{3}{2} c_v T.
\end{equation}
The Lagrangian can then be normalized by dividing by \(-\frac{3}{2} c_v\), and in terms of the \( b \)-parameters (2), it is expressed as:
\begin{equation}
-\frac{2}{3c_{v}}\mathcal{L} = \alpha \mathscr{A} + \beta \mathscr{T} + T, \quad \text{with} \quad \alpha = \frac{3 b_3}{2 c_v}, \quad \beta = -\frac{2 b_1}{3 c_v}.
\end{equation}
Alternatively, this is equivalent to setting \( c_v = -2/3 \) in Lagrangian \eqref{Lref}, which then takes the form:
\begin{equation}
\mathcal{L} = \left( -\frac{3}{2} + c_a \right) \mathscr{A} + \left( -\frac{2}{3} + c_t \right) \mathscr{T} + T.
\end{equation}
This expression can be rewritten as:
\begin{equation}\label{Lnor}
\mathcal{L} = c_a \mathscr{A} + c_t \mathscr{T} - \frac{2}{3} \mathscr{V}.
\end{equation}
Applying this normalization to the \( b \)-parameters, we obtain:
\begin{equation}\label{bpn}
b_1 = -\frac{2}{3}+ c_t, \quad b_2 = 3 c_t, \quad b_3 = \frac{2}{3} - \frac{4}{9} c_a.
\end{equation}
This implies that:
\begin{equation}\label{norma}
b_2 = 2 + 3 b_1.
\end{equation}
Note that an obvious consequence of choosing this normalization is that the definition of the \( b \)-parameters \eqref{bp} simplifies into a decoupled version, given by \eqref{bpn}. Furthermore, if we continue our study using this normalization, all results for the case \( c_v = 0 \) will be excluded. To obtain the most general results possible, we proceed with the Lagrangian \eqref{Lngr} and apply the normalization at the end, which allows us to separately identify the models described by Lagrangians \eqref{Lnon} and \eqref{Lnor}.
%
%
\section{Perturbative analysis}\label{pertuA}

Given the structure of the AFE and SFE, Eq.~\eqref{FEAV} and Eq.~\eqref{FESV} in vacuum, obtaining exact solutions is not straightforward when treating $\chi$ and $\psi$ as arbitrary functions. Since our main goal is to investigate possible black hole geometries, we introduce a perturbative approach in this section to facilitate such an analysis.

This method, previously applied in \cite{alanND2024} to study black hole in \( f(T) \) theories, is now adapted in the context of NGR. The core idea is to adopt a convenient coordinate gauge that simplifies the analysis. We then impose the necessary condition for the existence of a local horizon—namely, the vanishing of Eq.~\eqref{theAH}—to extract information about the behavior of the arbitrary functions involved.
Based on this, we propose an ansatz that enables us to reformulate the AFE and determine the conditions under which they are satisfied order by order. We subsequently analyze the SFE following a similar strategy. This procedure leads to distinct solution branches, which we classify into different cases to facilitate a later examination of the behaviour of the teleparallel scalars \eqref{TScalars} and \eqref{TS1}.

\subsection{Choice of coordinates}

Exploiting the freedom in the definition of the radial coordinate \( r \) can significantly simplify the FE. In the static case, where all arbitrary functions are time-independent, we may introduce a new coordinate system of the form \( t' = t \) and \( r' = r / A_{2} \). This transformation reduces the number of independent functions to four while preserving the diagonal structure of the tetrad \cite{robert2024}. Motivated by this, we adopt the coordinate choice
\begin{equation}\label{coor1}
A_{3} = r \, .
\end{equation}
This coordinate gauge is commonly employed in both GR and TG. In what follows, we analyze the AFE and  SFE using the coordinate condition \eqref{coor1}.

\subsection{Local horizon}\label{AP}

The condition for the existence of a local horizon at \( r = r_h \) is that the expansion scalar of outgoing null rays, $ \theta_{\sl \ell \sr}$, vanishes at that location, i.e., \( \theta_{\sl \ell \sr}(r_{h}) = 0 \) \cite{GH}. For the static, spherically symmetric case under consideration, and using the coordinate choice introduced in \eqref{coor1}, this coefficient takes the form
\begin{equation}\label{rho}
\theta_{\sl \ell \sr} = \frac{\sqrt{2}}{r}a_{2} \, , \quad a_{2} = \frac{1}{A_{2}} \, .
\end{equation}
Assuming $a_{2}=0$ at \( r = r_h \), we introduce a small parameter \( \epsilon > 0 \) and expand the radial coordinate as \( r = r_h + \epsilon \), with \( \epsilon \to 0 \). Under this assumption, we adopt the following ansatz:
\begin{equation}
a_2 = \epsilon^p (\alpha_1 + \alpha_2\, \epsilon),
\end{equation}
where \( p > 0 \). As a consequence, the function \( A_2 = 1/a_2 \) and $A_{1}$ take the form
\begin{equation}\label{AAe}
A_2 = \frac{\epsilon^{-p}}{\alpha_1 + \alpha_2\, \epsilon}, \quad A_1 = \epsilon^q (\beta_1 + \beta_2\, \epsilon),
\end{equation}
where \( A_1 \) is assumed to have a similar structure for consistency, with $q$ being an arbitrary constant. 

Since the arbitrary functions \( \chi \) and \( \psi \) play a key role in the FE, but the presence of a local horizon alone does not constrain their behaviour, we assume that these functions, defined in terms of \( r = r_h + \epsilon \), may follow a similar expansion structure to that of \eqref{AAe}. Thus, we consider:
\begin{equation}\label{CPe}
\chi = \epsilon^u (\chi_0 + \gamma_1 \, \epsilon), \quad \psi = \epsilon^v (\psi_0 + \gamma_2 \, \epsilon) \, ,
\end{equation}
where $u$ and $v$ are arbitrary constants. In the following, we examine whether the AFE can be satisfied under these assumptions, considering various combinations of the possible values for \( u \) and \( v \).
\subsection{Analysis of the AFE}\label{AFEp9}

Building on the previously introduced ansatz for the arbitrary functions \eqref{AAe} and \eqref{CPe}, we now reformulate the AFE \eqref{FEAV} in terms of the perturbation parameter \( \epsilon \), retaining terms up to first order. To streamline the analysis, we focus on identifying key values of the parameters \( u \) and \( v \), particularly in the case where \( u = 0 \) and \( v = 0 \). This leads to nine distinct cases to be investigated, categorized as follows:
\begin{center}
\begin{tabular}{l@{\hspace{1.5cm}}l@{\hspace{1.5cm}}l}
  1)\, $u > 0 \,\, \text{and} \,\, v > 0$ 
& 4)\, $u < 0 \,\, \text{and} \,\, v > 0$
& 7)\, $u = 0 \,\, \text{and} \,\, v > 0$ \\
  2)\, $u > 0 \,\, \text{and} \,\, v < 0$ 
& 5)\, $u < 0 \,\, \text{and} \,\, v < 0$ 
& 8)\, $u = 0 \,\, \text{and} \,\, v < 0$ \\
  3)\, $u > 0 \,\, \text{and} \,\, v = 0$ 
& 6)\, $u < 0 \,\, \text{and} \,\, v = 0$ 
& 9)\, $u = 0 \,\, \text{and} \,\, v = 0$
\end{tabular}
\end{center}

In the following, we examine whether the AFE are satisfied in each case, order by order. We focus on terms proportional to $\epsilon^w$ with exponents $w \leq 0$, as these represent the leading-order contributions. At each order, we analyze the corresponding coefficient and impose the necessary constraints to ensure that the term vanishes. When determining the dominant terms becomes an unbounded problem, we adopt an effective strategy by comparing terms with similar exponents and identifying which one is smaller, thereby avoiding the need for an exhaustive analysis of every possibility. Following this approach, the procedure is applied sequentially, starting from the leading-order terms and progressing to less significant contributions.

Since the SFE are expected to govern the behavior of the functions \( A_1 \) and \( A_2 \), our primary goal here is to identify the admissible values of the parameters associated with \( \chi \) and \( \psi \), where possible. Noting that case 9), \( u = 0 \) and \( v = 0 \), contains most of the relevant branches, we outline the method for this case as an example and summarize the results for the remaining cases in Section~\ref{AFEp}. Retaining only the leading-order terms, the AFE simplify to:

\begin{subequations}
\begin{flalign}
W_{\al12\ar} &= 
\frac{b_{\on} \, \gamma_{\tw} (p + q)}{2 \epsilon}
+ \frac{1}{2} b_{\on} \, \gamma_{\tw} \left( \frac{\alpha_{\tw}}{\alpha_{\on}} + \frac{\beta_{\tw}}{\beta_{\on}} + \frac{2}{r_{\h}} \right) + \frac{\epsilon^{1 - p} \, \gamma_{\tw} \, \cosh(\psi_{\n}) \left( b_{\on} \cos(\chi_{\n}) - b_{\tr} \gamma_{\on} r_{\h} \sin(\chi_{\n}) \right)}{\alpha_{\on} r_{\h}^2} & \nonumber \\
&\quad - \frac{\epsilon^{1 - p} \, \gamma_{\on} \left( b_{\tr} \gamma_{\on} r_{\h} \cos(\chi_{\n}) + b_{\on} \sin(\chi_{\n}) \right) \sinh(\psi_{\n})}{\alpha_{\on} r_{\h}^2}  + \frac{\epsilon^{-p} \left( b_{\on} \cos(\chi_{\n}) - b_{\tr} \gamma_{\on} r_{\h} \sin(\chi_{\n}) \right) \sinh(\psi_{\n})}{\alpha_{\on} r_{\h}^2} & \nonumber \\
&\quad + \frac{\epsilon^{1 - p} \left( - b_{\on} (2 \alpha_{\on} + \alpha_{\tw} r_{\h}) \cos(\chi_{\n}) + b_{\tr} \gamma_{\on} r_{\h} (\alpha_{\on} + \alpha_{\tw} r_{\h}) \sin(\chi_{\n}) \right) \sinh(\psi_{\n})}{\alpha_{\on}^2 r_{\h}^3} = 0 \, , &
\end{flalign}

\begin{flalign}
W_{\al34\ar} &= 
\frac{(b_{\on} + b_{\tr}) \, \gamma_{\on} (p + q)}{2 \epsilon}
+ \frac{(b_{\on} + b_{\tr}) \, \gamma_{\on} (2 \alpha_{\on} \beta_{\on} + \alpha_{\tw} \beta_{\on} r_{\h} + \alpha_{\on} \beta_{\tw} r_{\h})}{2 \alpha_{\on} \beta_{\on} r_{\h}}  - \frac{b_{\tr} \, \epsilon^{-1 - p} \, q \, \cosh(\psi_{\n}) \, \sin(\chi_{\n})}{\alpha_{\on} r_{\h}} & \nonumber \\
&\quad + \frac{\epsilon^{-p} \left( \alpha_{\on} \beta_{\on} (b_{\on} + b_{\tr}(-1 + q)) - \alpha_{\on} b_{\tr} \beta_{\tw} r_{\h} + \alpha_{\tw} b_{\tr} \beta_{\on} q r_{\h} \right) \cosh(\psi_{\n}) \, \sin(\chi_{\n})}{\alpha_{\on}^2 \beta_{\on} r_{\h}^2} & \nonumber \\
&\quad - \frac{b_{\tr} \, \epsilon^{-2p} \, \sin(2 \chi_{\n})}{\alpha_{\on}^2 r_{\h}^2} 
- \frac{2 b_{\tr} \, \epsilon^{1 - 2p} \left( \alpha_{\on} \gamma_{\on} r_{\h} \cos(2 \chi_{\n}) - (\alpha_{\on} + \alpha_{\tw} r_{\h}) \sin(2 \chi_{\n}) \right)}{\alpha_{\on}^3 r_{\h}^3} & \nonumber \\
&\quad - \frac{b_{\tr} \, \epsilon^{-p} \left( \gamma_{\on} q \cos(\chi_{\n}) \cosh(\psi_{\n}) + \gamma_{\tw} (1 + q) \sin(\chi_{\n}) \sinh(\psi_{\n}) \right)}{\alpha_{\on} r_{\h}} = 0 \, .&
\end{flalign}
\end{subequations}

\noindent The leading-order behavior is determined by the exponents \(-1\), \(-p\), \(-1 - p\), and \(-2p\). By considering the different regimes \(0 < p < 1\), \(p = 1\), and \(p > 1\), and imposing the vanishing of the corresponding coefficients of the dominant terms, we establish a systematic foundation for our analysis. In seeking branches that satisfy the AFE, we identify the cases summarized in Table~\ref{AFEcase9}, where a blank indicates that the parameter is unconstrained.

\begin{table}[!h]
\centering
\begin{tabular}{|c|c|c|c|c|c|c|c|c|c|c|c|c|c|c|c|}
\hline
\multicolumn{4}{|c|}{\text{NGR}} & \multicolumn{3}{c|}{$\chi$} & \multicolumn{3}{c|}{$\psi$} & \multicolumn{3}{c|}{$A_1$}  & \multicolumn{3}{c|}{$A_2$} \\
\cline{1-16}
\# & $b_{\on}$ & $b_{\tw}$ & $b_{\tr}$ & $u$ & $\chi_{\n}$ & $\gamma_{\on}$ & $v$ & $\psi_{\n}$ & $\gamma_{\tw}$ & $q$ & $\beta_{\on}$ & $\beta_{\tw}$ & $p$ & $\alpha_{\on}$ & $\alpha_{\tw}$ \\
\hline
9.1 & 0 &   & 0         & 0 &   &   & 0 &   &   &   &   &   & $(0,\infty)$ & & \\
9.2 & 0 &   &           & 0 & $n\pi$ & 0 & 0 &   &   &   &   &   & $(0,\infty)$ & & \\
9.3 &   &   &           & 0 & $n\pi$ & 0 & 0 & 0 & 0 &   &   &   & $(0,\infty)$ & & \\
9.4 &   &   & $-b_{\on}$ & 0 & $n\pi$ &   & 0 & 0 & 0 & 0 &   &   & $(0,\tfrac{1}{2})$ & & \\
9.5 &   &   & 0         & 0 & $n\pi$ &   & 0 & 0 & 0 & $-p$ &   & $a\beta_{\on}$ & $(0,\infty)$ & & $-\left(\dfrac{2}{r_{\h}} + a\right)\alpha_{\on}$ \\[11pt]
9.6 &   &   &           & 0 & $n\pi$ & 0 & 0 & 0 &   & $-p$ &   & $a\beta_{\on}$ & $(0,1)$ & & $-\left(\dfrac{2}{r_{\h}} + a\right)\alpha_{\on}$ \\[11pt]
\hline
\end{tabular}
\caption{Parameter values that satisfy the AFE for the case \( u = 0 \) and \( v = 0 \), where $a=\beta_{\tw}/\beta_{\on}$. }
\label{AFEcase9}
\end{table}
\subsection{Analysis of the SFE}\label{SFEp9}

In the previous analysis, we identified several branches that satisfy the AFE. However, a thorough examination of the SFE is still required. To address this, we build upon the earlier results and perform a perturbative analysis of the SFE. Using the previously introduced ansatz for the arbitrary functions in Eqs.~\eqref{AAe} and \eqref{CPe}, we reformulate the SFE in terms of the perturbation parameter $\epsilon$ for the following distinct cases:

\begin{center}
\begin{tabular}{l@{\hspace{1.5cm}}l@{\hspace{1.5cm}}l}
  1)\, $0<u\leq 1 \,\, \text{and} \,\, 0<v\leq 1$ 
& 4)\, $u\leq -1 \,\, \text{and} \,\, 0<v\leq 1$
& 7)\, $u = 0 \,\, \text{and} \,\, 0<v\leq 1$ \\
  2)\, $0<u\leq 1 \,\, \text{and} \,\, v\leq -1$ 
& 5)\, $u\leq -1 \,\, \text{and} \,\, v\leq -1$ 
& 8)\, $u = 0 \,\, \text{and} \,\, v\leq -1$ \\
  3)\, $0<u\leq 1 \,\, \text{and} \,\, v = 0$ 
& 6)\, $u\leq -1 \,\, \text{and} \,\, v = 0$ 
& 9)\, $u = 0 \,\, \text{and} \,\, v = 0$
\end{tabular}
\end{center}

Since the values of \( b_{1} \), \( b_{2} \), and \( b_{3} \) determine the NGR models under consideration, we seek solutions while avoiding fixing these constants whenever possible. If solving the SFE requires imposing strong conditions that lead to the trivial solution \( b_{1} = b_{2} = b_{3} = 0 \), we discard that branch. Likewise, we discard any branch that leads to inconsistencies at any order in \( \epsilon \) among the terms we have retained, as well as any branch that is evidently a particular case of another within the same analysis. We now proceed with the example for case 9), \( u = 0 \) and \( v = 0 \), and outline the results for the remaining cases in Appendix~\ref{SFEp}. Retaining only the leading-order terms, the SFE simplify to:

\begin{subequations}
\begin{flalign}
W_{\sl 11\sr} &= \frac{\epsilon^{-2p} \left( -4 b_{\on} + b_{\tw} - 2 b_{\tr} + 2 b_{\tr} \cos(2 \chi_{\n}) \right)}{2 \alpha_{\on}^2 r_{\h}^2} + \frac{-\alpha_{\on} b_{\tw} \beta_{\on} p + \alpha_{\tw} b_{\on} \beta_{\on} q r_{\h} + \alpha_{\on} b_{\on} (p + q)(2 \beta_{\on} + \beta_{\tw} r_{\h})}{\alpha_{\on} \beta_{\on} \epsilon r_{\h}} && \nonumber \\
&\quad +\frac{b_{\on} q (-2 + 2p + q)}{2 \epsilon^2}  + \frac{2 \epsilon^{-p} \cosh(\psi_{\n}) \left( -b_{\on} \cos(\chi_{\n}) + b_{\tr} \gamma_{\on} r_{\h} \sin(\chi_{\n}) \right)}{\alpha_{\on} r_{\h}^2} = 0 \, , &
\end{flalign}
\begin{flalign}
W_{\sl 22\sr} &= -\frac{b_{\on} q^2}{2 \epsilon^2} + \frac{q \left( b_{\tw} \beta_{\on} - b_{\on}(2 \beta_{\on} + \beta_{\tw} r_{\h}) \right)}{\beta_{\on} \epsilon r_{\h}}  + \frac{\epsilon^{-2p} \left( 4 b_{\on} - b_{\tw} + 2 b_{\tr} - 2 b_{\tr} \cos(2 \chi_{\n}) \right)}{2 \alpha_{\on}^2 r_{\h}^2} = 0 \, , &
\end{flalign}
\begin{flalign}
W_{\sl 33\sr} &= \frac{q \left( b_{\tw} (-1 + p + q) - b_{\on} (-2 + 2p + q) \right)}{2 \epsilon^2} && \nonumber \\
&\quad + \frac{\alpha_{\tw} (-2 b_{\on} + b_{\tw}) \beta_{\on} q r_{\h} + \alpha_{\on} \left( b_{\tw} \beta_{\on} (p + q) + b_{\tw} \beta_{\tw} (p + 2q) r_{\h} - 2 b_{\on} (p + q)(2 \beta_{\on} + \beta_{\tw} r_{\h}) \right)}{2 \alpha_{\on} \beta_{\on} \epsilon r_{\h}} && \nonumber \\
&\quad - \frac{b_{\on} \epsilon^{-1 - p} q \cos(\chi_{\n}) \cosh(\psi_{\n})}{\alpha_{\on} r_{\h}} + \frac{\epsilon^{-p} \gamma_{\on} (b_{\on} - b_{\tr} + b_{\on} q) \cosh(\psi_{\n}) \sin(\chi_{\n})}{\alpha_{\on} r_{\h}} && \nonumber \\
&\quad + \frac{b_{\on} \epsilon^{-p} \cos(\chi_{\n}) \left[ \left( \alpha_{\on} \beta_{\on} q - \alpha_{\on} \beta_{\tw} r_{\h} + \alpha_{\tw} \beta_{\on} q r_{\h} \right) \cosh(\psi_{\n}) - \alpha_{\on} \beta_{\on} \gamma_{\tw} (1 + q) r_{\h} \sinh(\psi_{\n}) \right]}{\alpha_{\on}^2 \beta_{\on} r_{\h}^2} = 0 \, . &
\end{flalign}
\end{subequations}

\noindent Since the leading-order terms are governed by the exponents \(-2\), \(-2p\), and \(-1 - p\), our analysis considers the different regimes \(0 < p < 1\), \(p = 1\), and \(p > 1\), and is guided by the requirement that the coefficients of the dominant terms vanish. Introducing the parameter
\begin{equation}
\bar{\beta}_{\on} = -\dfrac{\beta_{\on}(\alpha_{\tw} r_{\h}^{3/2} + 2\delta)}{3 r_{\h} \delta} \, ,
\end{equation}
we identify the branches that simultaneously satisfy both the AFE and the SFE, which are summarized in Table~\ref{FEcase9}, where a blank indicates that the parameter is unconstrained.

\begin{table}[!h]
\centering
\begin{tabular}{|c|c|c|c|c|c|c|c|c|c|c|c|c|c|}
\hline
\multicolumn{4}{|c|}{\text{NGR}} & \multicolumn{2}{c|}{$\chi$} & \multicolumn{2}{c|}{$\psi$} & \multicolumn{3}{c|}{$A_1$} & \multicolumn{3}{c|}{$A_2$} \\
\cline{1-14}
\# & $b_{\on}$ & $b_{\tw}$ & $b_{\tr}$ & $\chi_{\n}$ & $\gamma_{\on}$ & $\psi_{\n}$ & $\gamma_{\tw}$ & $q$ & $\beta_{\on}$ & $\beta_{\tw}$ & $p$ & $\alpha_{\on}$ & $\alpha_{\tw}$ \\
\hline
9.1.1 & $0$ &  & $0$ &  &  &  &  & $\frac{1}{2}$ &  & $\bar{\beta}_{\on}$ & $\frac{1}{2}$ & $\dfrac{\delta}{\sqrt{r_{\h}}}$ & \\[11pt]
9.2.1 & $0$ &  &  & $n\pi$ & $0$ &  &  & $\frac{1}{2}$ &  & $\bar{\beta}_{\on}$ & $\frac{1}{2}$ & $\dfrac{\delta}{\sqrt{r_{\h}}}$ & \\[11pt]
9.3.1 &  & $0$ &  & $n\pi$ & $0$ & $0$ & $0$ & $-\dfrac{2\delta}{\alpha_{\on} r_{\h}}$ &  & $-\dfrac{2\beta_{\on}}{r_{\h}}$ & $1$ &  & $-\dfrac{\alpha_{\on}}{r_{\h}}$ \\[11pt]
9.3.2 &  & $3 b_{\on}$ &  & $n\pi$ & $0$ & $0$ & $0$ & $\dfrac{\delta}{\alpha_{\on} r_{\h}}$ &  & $\dfrac{\beta_{\on}}{r_{\h}}$ & $1$ &  & $-\dfrac{\alpha_{\on}}{r_{\h}}$ \\[11pt]
9.3.3 &  & $4 b_{\on}$ &  & $n\pi$ & $0$ & $0$ & $0$ & $0$ &  & $0$ & $1$ & $-\dfrac{\delta}{r_{\h}}$ & \\[11pt]
9.3.4 &  & $4 b_{\on}$ &  & $n\pi$ & $0$ & $0$ & $0$ & $0$ &  & $\dfrac{4\beta_{\on}}{r_{\h}}$ & $1$ & $\dfrac{\delta}{r_{\h}}$ & \\[11pt]
\hline
\end{tabular}
\caption{Parameter values that satisfy both the AFE and SFE for the case \( u =0 \) and \( v=0 \). }
\label{FEcase9}
\end{table}

\subsection{Classification of cases}

After analyzing the conditions under which both the AFE and the SFE are satisfied order by order, we observe that many branches overlap or correspond to the same branch under different conditions on the parameters $u$ and $v$.

Moreover, some branches are identified as qualitatively equivalent or as particular cases of more general ones. Therefore, we retain only the most general branches and consolidate the distinct cases in Table~\ref{casesIDNorma}, where a blank indicates that the parameter is unconstrained. 

Since only Case 4 cannot be normalized (\( c_{v} = 0 \)), we apply the normalization from \eqref{norma} to the remaining cases (\( c_{v} \neq 0 \)).
The analysis presented in \citep{thesis} showed that, in addition to Case 4, two special subcases—referred to as Cases 4.2 and 4.3 in \citep{thesis}, also could not be incorporated into the normalized version of the theory \eqref{Lnor}.
These cases are characterized by the condition $c_{v} = 0$. In contrast, Case 1 corresponds to TEGR, Case 2 represents the 1P H\&S model, and Cases 3, 5.1, and 5.2 are 1P (after normalization) special cases that cannot be reduced to TEGR in any limit.
It was shown in \citep{thesis} that in Cases 4.2 and 4.3, the leading coefficients $\psi_{\n}$ and $\chi_{\n}$ vanish. Since $v > 0$ and $u > 0$ in these cases, the remaining terms $\gamma_{\on}$ and $\gamma_{\tw}$ in the AFE appeared multiplied by positive powers of $\epsilon$, allowing them to be neglected without imposing further conditions. To assess whether $\gamma_{\on}$ and $\gamma_{\tw}$ were genuinely free parameters, we redefined $\psi$ and $\chi$ assuming $\psi_{\n} = 0$ and $\chi_{\n} = 0$, and reapplied the algorithm to derive explicit forms for both functions.

In Case 4.2, setting $\psi_{\n} = 0$ yielded a modified form of $\psi$, prompting a reformulation of the AFE in terms of $\epsilon$, focusing on leading-order contributions with $u = 0$ and $v > 0$. The equation $W_{\al34\ar}$ vanished when $\chi = n\pi$, while $W_{\al12\ar}$ required either $b_1 = 0$ or $\gamma_{\tw} = 0$; the latter being chosen to avoid overlap with Case 2. The corresponding SFE vanished only if $b_2 = 0$ or $b_2 = 3b_1$, aligning with subcases of Cases 3 and 4. In Case 4.3, assuming $\chi_{\n} = 0$, a similar analysis was carried out with $u > 0$, leading to simplified AFE. Here, $W_{\al12\ar}$ vanished when $\psi = 0$, and $W_{\al34\ar}$ required either $b_3 = -b_1$ or $\gamma_{\on} = 0$. The SFE again imposed the same conditions, reducing the scenario to known subcases of Cases 3 and 4. Consequently, both Case 4.2 and Case 4.3 were not independent and were therefore discarded. 

\subsection{Analysis of teleparallel scalars}

In the previous analysis, five independent cases were identified, each corresponding to specific parameter values that satisfy the AFE and SFE at the dominant perturbative order. Notably, in Case 1, both \( \chi \) and \( \psi \) remain completely arbitrary, while in Case 2, \( \chi = n\pi \) with \( \psi \) still arbitrary. In contrast, Cases 3 through 5 fully determine \( \chi \) and \( \psi \), fixing them to the constant values \( \chi = n\pi \) and \( \psi = 0 \). 

\begin{table}[!h]
\centering
\renewcommand{\arraystretch}{1}
\rotatebox{90}{
\begin{tabular}{|c|c|c|c|c|c|c|c|c|c|c|c|c|c|c|c|c|}
\hline
\multicolumn{5}{|c|}{\text{NGR}} & \multicolumn{3}{c|}{$\chi$} & \multicolumn{3}{c|}{$\psi$} & \multicolumn{3}{c|}{$A_1$} & \multicolumn{3}{c|}{$A_2$} \\
\cline{1-17}
\# & $b_{\on}$ & $b_{\tw}$ & $b_{\tr}$ & Lagrangian & $u$ & $\chi_{\n}$ & $\gamma_{\on}$ & $v$ & $\psi_{\n}$ & $\gamma_{\tw}$ & $q$ & $\beta_{\on}$ & $\beta_{\tw}$ & $p$ & $\alpha_{\on}$ & $\alpha_{\tw}$ \\
\hline
1 & $0$ & $2$ & $0$ & $T$ &  &  &  &  &  &  & $\frac{1}{2}$ &  & $\bar{\beta}_{\on}$ & $\frac{1}{2}$ & $\dfrac{\delta}{\sqrt{r_{\h}}}$ & \\[11pt]
2 & $0$ & $2$ & $\dfrac{2}{3} - \dfrac{4}{9} c_{a}$ & $c_{a} \mathscr{A} + \dfrac{2}{3} \mathscr{T} - \dfrac{2}{3} \mathscr{V}$ & $0$ & $n\pi$ & $0$ &  &  &  & $\frac{1}{2}$ &  & $\bar{\beta}_{\on}$ & $\frac{1}{2}$ & $\dfrac{\delta}{\sqrt{r_{\h}}}$ & \\[11pt]
3 & $-\frac{2}{3}$ & $0$ & $\dfrac{2}{3} - \dfrac{4}{9} c_{a}$ & $c_{a} \mathscr{A} - \dfrac{2}{3} \mathscr{V}$ & $0$ & $n\pi$ & $0$ &  & $0$ & $0$ & $-\dfrac{2\delta}{\alpha_{\on} r_{\h}}$ &  & $-\dfrac{2\beta_{\on}}{r_{\h}}$ & $1$ &  & $-\dfrac{\alpha_{\on}}{r_{\h}}$ \\[11pt]
4 & $c_{t}$ & $3 c_{t}$ & $-\dfrac{4}{9} c_{a}$ & $c_{a} \mathscr{A} + c_{t} \mathscr{T}$ & $0$ & $n\pi$ & $0$ &  & $0$ & $0$ & $\dfrac{\delta}{\alpha_{\on} r_{\h}}$ &  & $\dfrac{\beta_{\on}}{r_{\h}}$ & $1$ &  & $-\dfrac{\alpha_{\on}}{r_{\h}}$ \\[11pt]
5.1 & $2$ & $8$ & $\dfrac{2}{3} - \dfrac{4}{9} c_{a}$ & $c_{a} \mathscr{A} + \dfrac{8}{3} \mathscr{T} - \dfrac{2}{3} \mathscr{V}$ & $0$ & $n\pi$ & $0$ &  & $0$ & $0$ & $0$ &  & $0$ & $1$ & $-\dfrac{\delta}{r_{\h}}$ & \\[11pt]
5.2 & $2$ & $8$ & $\dfrac{2}{3} - \dfrac{4}{9} c_{a}$ & $c_{a} \mathscr{A} + \dfrac{8}{3} \mathscr{T} - \dfrac{2}{3} \mathscr{V}$ & $0$ & $n\pi$ & $0$ &  & $0$ & $0$ & $0$ &  & $\dfrac{4\beta_{\on}}{r_{\h}}$ & $1$ & $\dfrac{\delta}{r_{\h}}$ & \\[11pt]
\hline
\end{tabular}
}
\caption{Normalized NGR models (except Case 4) and parameter values satisfying both the AFE and the SFE.}
\label{casesIDNorma}
\end{table}

\begin{table}[H]
\centering
\renewcommand{\arraystretch}{1}
\rotatebox{90}{
\begin{tabular}{|c|c|c|c|c|c|c|c|c|c|c|c|c|c|c|c|c|c|c|}
\hline
\multicolumn{5}{|c|}{\text{NGR}} & \multicolumn{2}{c|}{$\chi$} & $\psi_{\n}$ & \multicolumn{3}{c|}{$A_1$} & \multicolumn{3}{c|}{$A_2$} & \multicolumn{4}{c|}{$\epsilon \to 0$} \\
\cline{1-18}
\# & $b_{\on}$ & $b_{\tw}$ & $b_{\tr}$ & Lagrangian & $\chi_{\n}$ & $\gamma_{\on}$ &  & $q$ & $\beta_{\on}$ & $\beta_{\tw}$ & $p$ & $\alpha_{\on}$ & $\alpha_{\tw}$   & $\mathscr{T}$ & $\mathscr{V}$ & $\mathscr{A}$ & $T$ \\
\hline
1.1 & $0$ & $2$ & $0$ & $T$ & $\dfrac{\pi}{2}$ &  &  & $\dfrac{1}{2}$ &  & $\bar{\beta}_{\on}$ & $\dfrac{1}{2}$ & $\dfrac{\delta}{\sqrt{r_{\h}}}$ &  & $\infty$ & $\infty$ & $-\dfrac{16}{9 r_{\h}^2}$ & $-\dfrac{4}{r_{\h}^2}$ \\[11pt]
1.2 & $0$ & $2$ & $0$ & $T$ & $n\pi$ &  &  & $\dfrac{1}{2}$ &  & $\bar{\beta}_{\on}$ & $\dfrac{1}{2}$ & $\dfrac{\delta}{\sqrt{r_{\h}}}$ &  & $\infty$ & $\infty$ & $0$ & $\infty$ \\[11pt]
2 & $0$ & $2$ &  & $c_{a} \mathscr{A} + \dfrac{2}{3} \mathscr{T} - \dfrac{2}{3} \mathscr{V}$ & $n\pi$ & $0$ &  & $\dfrac{1}{2}$ &  & $\bar{\beta}_{\on}$ & $\dfrac{1}{2}$ & $\dfrac{\delta}{\sqrt{r_{\h}}}$ &  & $\infty$ & $\infty$ & $0$ & $\infty$ \\[11pt]
3 & $-\dfrac{2}{3}$ & $0$ &  & $c_{a} \mathscr{A} - \dfrac{2}{3} \mathscr{V}$ & $n\pi$ & $0$ & $0$ & $-\dfrac{2\delta}{\alpha_{\on} r_{\h}}$ &  & $-\dfrac{2\beta_{\on}}{r_{\h}}$ & $1$ &  & $-\dfrac{\alpha_{\on}}{r_{\h}}$ & $\dfrac{9}{r_{\h}^2}$ & $0$ & $0$ & $\dfrac{6}{r_{\h}^2}$ \\[11pt]
4 & $c_{t}$ & $3 c_{t}$ &  & $c_{a} \mathscr{A} + c_{t} \mathscr{T}$ & $n\pi$ & $0$ & $0$ & $\dfrac{\delta}{\alpha_{\on} r_{\h}}$ &  & $\dfrac{\beta_{\on}}{r_{\h}}$ & $1$ &  & $-\dfrac{\alpha_{\on}}{r_{\h}}$ & $0$ & $\dfrac{9}{r_{\h}^2}$ & $0$ & $-\dfrac{6}{r_{\h}^2}$ \\[11pt]
5.1 & $2$ & $8$ &  & $c_{a} \mathscr{A} + \dfrac{8}{3} \mathscr{T} - \dfrac{2}{3} \mathscr{V}$ & $n\pi$ & $0$ & $0$ & $0$ &  & $0$ & $1$ & $-\dfrac{\delta}{r_{\h}}$ &  & $\dfrac{1}{r_{\h}^2}$ & $\dfrac{4}{r_{\h}^2}$ & $0$ & $-\dfrac{2}{r_{\h}^2}$ \\[11pt]
5.2 & $2$ & $8$ &  & $c_{a} \mathscr{A} + \dfrac{8}{3} \mathscr{T} - \dfrac{2}{3} \mathscr{V}$ & $n\pi$ & $0$ & $0$ & $0$ &  & $\dfrac{4\beta_{\on}}{r_{\h}}$ & $1$ & $\dfrac{\delta}{r_{\h}}$ &  & $\dfrac{1}{r_{\h}^2}$ & $\dfrac{4}{r_{\h}^2}$ & $0$ & $-\dfrac{2}{r_{\h}^2}$ \\[11pt]
\hline
\end{tabular}
}\caption{Normalized NGR models (excluding Case 4), parameter values satisfying both the AFE and SFE, and the behavior of the teleparallel scalar as $\epsilon \to 0^{+}$.}
\label{sum}
\end{table}

We now focus on the teleparallel scalars defined in Eqs.~\eqref{TorS2} and \eqref{TdeS}. Using Eqs.~\eqref{AAe} and \eqref{CPe}, these invariants can be expressed in terms of the perturbation parameter \( \epsilon \), taking \( u = 0 \) and \( v = 0 \) as a representative example. Retaining only the leading-order terms, the teleparallel scalars in Eq.~\eqref{TScalars} are given by:
\begin{subequations}
\begin{alignat}{2}\nonumber
\mathscr{V} = &\, \alpha_{\on}^{2} \epsilon^{-2 + 2p} q^{2} + \frac{4 \alpha_{\on} \epsilon^{-1 + p} q \cos(\chi_{\n}) \cosh(\psi_{\n})}{r_{\h}} \\ \label{vl}
&+ \frac{4 \cos^{2}(\chi_{\n})}{r_{\h}^{2}}  + \frac{2 \alpha_{\on} \epsilon^{-1 + 2p} q \left(2 \alpha_{\on} \beta_{\on} + \alpha_{\on} \beta_{\tw} r_{\h} + \alpha_{\tw} \beta_{\on} q r_{\h}\right)}{\beta_{\on} r_{\h}}\,,
\\ \label{al}
\mathscr{A} = &\, -\frac{16 \sin^2(\chi_{\n})}{9 r_{\h}^2} \,,\\ \nonumber 
\mathscr{T} = &\, \alpha_{\on}^{2} \epsilon^{-2 + 2p} q^{2}  - \frac{2 \alpha_{\on} \epsilon^{-1 + p} q \cos(\chi_{\n}) \cosh(\psi_{\n})}{r_{\h}} \\ \label{tl}
&+ \frac{1}{r_{\h}^{2}} + \frac{2 \alpha_{\on} \epsilon^{-1 + 2p} q \left(-\alpha_{\on} \beta_{\on} + \alpha_{\on} \beta_{\tw} r_{\h} + \alpha_{\tw} \beta_{\on} q r_{\h}\right)}{\beta_{\on} r_{\h}} \,.
\end{alignat}
\end{subequations}
We observe that \( \mathscr{A} \) remains finite in the limit \( \epsilon \to 0^{+} \); however, the other invariants remain well behaved only if \( p \geq 1 \) or \( q = 0 \). The same behaviour is seen for the torsion scalar given in Eq.~\eqref{TS1}, which in this case takes the form:
\begin{equation}
T = -\frac{2}{r_{\h}^{2}} - \frac{4 \alpha_{\on}^{2}q}{r_{\h}}\, \epsilon^{-1 + 2p} - \frac{4 \alpha_{\on}  q \cos(\chi_{\n}) \cosh(\psi_{\n})}{r_{\h}}\, \epsilon^{-1 + p} \, .
\end{equation}
\noindent Therefore, for \( p < 1 \) and \( q \neq 0 \), the scalars \( \mathscr{T} \) and \( \mathscr{V} \) inevitably diverge as \( \epsilon \to 0^{+} \). However, this is not necessarily the case for the torsion scalar \( T \). For instance, when \( p = 1/2 \), whether \( T \) diverges depends on the value of \( \chi_{\n} \); in particular, if \( \chi_{\n} = (n \pm 1/2)\pi \), the torsion scalar remains finite, otherwise it diverges.

Using the corresponding parameter values for the functions \( A_{1} \) and \( A_{2} \) in each of the identified cases listed in Table~\ref{casesIDNorma}, we can analyse the behaviour of the scalars as \( \epsilon \to 0^{+} \) for each model, as summarized in Table~\ref{sum}, where a blank indicates that the parameter is unconstrained.

%
%
\section{Discussion}

A summary of the branches that satisfy both the AFE and SFE is presented in Table~\ref{sum}. We now discuss the key features of these results. As a reference, we use Table~\ref{GFC}, where DNPS-2 stands for ``Does Not Propagate Spin-2''. This table summarizes the different types of NGR analyzed in \citep{them}, along with the necessary conditions for the theory to be ghost-free.

\begin{table}[!h]
\centering
\renewcommand{\arraystretch}{1.4}
\begin{tabular}{|c|c|c|}
\hline
\text{Theory} & \text{Parameter space} & \text{Ghost-free condition} \\
\hline
Type 1 & Generic & Impossible \\
Type 2 & $b_{\on} = 0$ & $b_{\tr} > 0 \quad \text{and} \quad b_{\tw} > 0$ \\
Type 3 & $b_{\tr} = -b_{\on}$ & $0 < b_{\tw} < 3 b_{\on} \quad \text{or} \quad 3 b_{\on} < 0 < b_{\tw}$ \\
Type 4 & $b_{\tw} = 0$ & DNPS-2 \\
Type 5 & $b_{\tw} = 3 b_{\on}$ & $b_{\tw} > 0 \quad \text{and} \quad b_{\on} > -b_{\tr}$ \\
Type 6 & $b_{\on} = 0 \quad \text{and} \quad b_{\tr} = 0$ & $b_{\tw} > 0$ \\
Type 7 & $b_{\tw} = 0 \quad \text{and} \quad b_{\tr} = -b_{\on}$ & DNPS-2 \\
Type 8 & $b_{\tw} = 3 b_{\on} \quad \text{and} \quad b_{\tr} = -b_{\on}$ & $b_{\tw} > 0$ \\
Type 9 & $b_{\on} = 0 \quad \text{and} \quad b_{\tw} = 0$ & DNPS-2 \\
\hline
\end{tabular}
\caption{Ghost-free condition for the NGR parameters. Adapted from \citep{them}.}
\label{GFC}
\end{table}
With the exception of the two branches that correspond essentially to the TEGR case (1 and 2), all other entries in Table~\ref{casesIDNorma} satisfy \( \psi = 0 \) and \( \chi = n\pi \). In several cases (specifically Types 4, 7, and 9 \citep{them}), a spin-2 field fails to propagate in general, not just in the spherically symmetric case, indicating that these theories cannot support gravitational wave solutions.
 In general, we have \( c_v \neq 0 \), and a useful normalization condition can be imposed: \( b_2 = 2 + 3b_1 \), though this relation does not apply to case 4.

\begin{itemize}
\item Case 1 (including 1.1 and 1.2) corresponds to TEGR, a theory with a well-defined Newtonian limit and consistency with solar system observations. It is classified as Type 6 in~\citep{them}, and its ghost-free condition requires \( b_2 > 0 \), which is satisfied here since \( b_2 = 2 \).
In Table~\ref{sum}, we present two different examples, cases 1.1 and 1.2, in which the scalars \( \mathscr{V} \) and  \( \mathscr{T} \) exhibit unavoidable divergences at the local horizon, indicating a geometric singularity. However, this behavior does not extend to the torsion scalar \( T \) in general.
For case 1.1, \( T \) remains finite at \( r_{\h} \), suggesting that the region \( r = r_{\h} \) and its interior (excluding \( r = 0 \)) are part of the spacetime manifold from the perspective of the theory’s action. In contrast, for case 1.2, \( T \) diverges at \( r_{\h} \), implying that the horizon (and its interior) must be excluded from the manifold domain.
Since in both cases the geometry remains divergent at the local horizon, the theory is ultimately unable to describe a well-defined black hole configuration.

\item Case 2 corresponds to the 1P Hayashi and Shirafuji model (1P-H\&S) at the level of the Lagrangian. However, since \( \chi = n\pi \) causes the parameter \( b_3 \) to decouple from the FE, the theory effectively reduces to TEGR. It is classified as Type 2 in~\citep{them}, with the ghost-free conditions requiring \( b_2 > 0 \) and \( b_3 > 0 \). The condition \( b_2 > 0 \) is satisfied with \( b_2 = 2 \), while \( b_3 > 0 \) holds when \( c_a < 3/2 \). As in Case 1, this theory is unable to describe a well-defined black hole configuration.

\item Case 3 corresponds to a 1P model at the level of the Lagrangian. However, due to the choice \( \chi = n\pi \), the parameter \( b_3 \) decouples from the FE, effectively reducing the theory to a 0P model that is distinct from TEGR and lacks a Newtonian limit. Moreover, with \( b_2 = 0 \), this case is classified as Type 4 in~\citep{them}, which does not support a propagating spin-2 field, implying the absence of gravitational waves. In this case, the geometry is well behaved at the local horizon, and the region \( r = r_{\h} \) and its interior (excluding \( r = 0 \)) are included in the manifold.

\item Case 4 arises when \( c_v = 0 \), or equivalently \( b_2 = 3b_1 \), which prevents the model from being included in the normalization. At the Lagrangian level, it corresponds to a 2P model. However, since \( \chi = n\pi \) causes \( b_3 \) to decouple from the FE, the theory effectively reduces to a 1P model. The only remaining parameter in the FE is \( c_t \), which appears as an overall factor. Provided \( c_t \neq 0 \), a normalization can be applied, reducing the system to a 0P model distinct from TEGR and therefore lacking a Newtonian limit.
This theory is classified as Type 5 in~\citep{them}, with ghost-free conditions requiring \( b_2 > 0 \) and \( b_1 > -b_3 \). Since \( b_1 \), \( b_2 \), and \( b_3 \) depend on \( c_t \) and \( c_a \), these conditions translate into the constraints \( c_t > \tfrac{4}{9}c_a \) and \( c_t > 0 \). With an appropriate choice of parameters, the ghost-free conditions can be satisfied.
In this case, the geometry is well behaved at the local horizon, and the region \( r = r_{\h} \) and its interior (excluding \( r = 0 \)) are included in the manifold.

\item Cases 5 (including 5.1 and 5.2) describe a 1P model at the level of the Lagrangian. However, as in previous cases, the condition \( \chi = n\pi \) causes \( b_3 \) to decouple from the FE, effectively reducing the theory to a 0P model that is distinct from TEGR and lacks a Newtonian limit.
For the theory considered here, it is possible to choose \( b_1 \), \( b_2 \), and \( b_3 \) such that it falls under Type 3 in~\citep{them}, where \( b_3 = -b_1 \). For instance, choosing \( c_a \) such that \( b_3 = -2 \), and taking \( b_1 = 2 \) and \( b_2 = 8 \), the ghost-free conditions, \( 0 < b_2 < 3b_1 \) or \( 3b_1 < 0 < b_2 \), are clearly not satisfied. Alternatively, if this theory is interpreted as a special case of Type 1 in~\citep{them}, then ghost-free modes are inherently excluded.
Consequently, these models cannot avoid ghost instabilities. Nevertheless, the geometry is well behaved at the local horizon, and the region \( r = r_{\h} \) and its interior (excluding \( r = 0 \)) are included in the  manifold.

\end{itemize}

\section{Summary and conclusions}
TEGR and the 1P-H\&S family were found to be incapable of describing black hole configurations. All remaining cases (2--5) correspond to 0P models that could, in principle, be well behaved at the local horizon. However, since these models exhibit certain unphysical features, no further analysis was pursued; e.g., the lower-order conditions from the AFE and SFE were not explored. These results complement those obtained in \citep{alanND2024} for the cases of \( F(T) \) gravity and TEGR. Nevertheless, further work is needed to fully understand the static, spherically symmetric, vacuum teleparallel geometries described by TEGR. 

For example, it remains to be explored whether the singular behavior of torsion scalars at the local horizon can be eliminated by appropriately choosing the functions \( \chi \) and \( \psi \), raising questions about their physical or geometrical interpretation and the conditions that might constrain them. It is also important to determine whether a well-behaved tetrad exists at the local horizon that avoids such singularities within TEGR.

%
%

\section*{Acknowledgement}

D.F.L. acknowledges support from the Department of Mathematics and Statistics at Dalhousie University, Canada. A.A.C. and R.J.vdH. are supported by the Natural Sciences and Engineering Research Council of Canada (NSERC). The authors thank N.T. Layden, J. Tot for valuable discussions and A. Golovnev for helpful correspondence.

%
%

\newpage

%
%

\appendix

\begin{center}
\vskip 0.7cm
{\Large \bf Appendix \\ }
\vskip 0.7cm
\end{center}

\renewcommand{\theequation}{\Alph{section}.\arabic{equation}}
\setcounter{equation}{0}

\section{Results of the AFE perturbative analysis}\label{AFEp}
In this section, we summarize the results of the perturbative analysis of the AFE. For a complete and detailed treatment, refer to \citep{thesis}. Note that all cases are presented here except for the case $u=v=0$, which was outlined in Section~\ref{AFEp9}.

\subsection{$\boldsymbol{u > 0 \,\, \text{and} \,\, v > 0}$}

To account for contributions from all leading terms, and given that \( p > 0 \), we restrict our analysis to the range \( 0 < u \leq 1 \) and \( 0 < v \leq 1 \). The leading-order behavior is determined by the exponents \( -2 + v \), \( -2 + u \), and \( 1 - p \). To identify the branches that satisfy the AFE across all possible configurations—including cases where multiple terms contribute simultaneously (i.e., when some of the exponents are equal)—we impose that the corresponding coefficients vanish. These possibilities are organized in Table~\ref{AFEcase1}.

\begin{table}[!h]
\centering
\renewcommand{\arraystretch}{1}
\begin{tabular}{|c|c|c|c|c|c|c|c|c|c|c|c|c|c|c|c|}
\hline
\multicolumn{4}{|c|}{\text{NGR}} & \multicolumn{3}{c|}{$\chi$} & \multicolumn{3}{c|}{$\psi$} & \multicolumn{3}{c|}{$A_1$} & \multicolumn{3}{c|}{$A_2$} \\
\cline{1-16}
\# & $b_{\on}$ & $b_{\tw}$ & $b_{\tr}$ & $u$ & $\chi_{\n}$ & $\gamma_{\on}$ & $v$ & $\psi_{\n}$ & $\gamma_{\tw}$ & $q$ & $\beta_{\on}$ & $\beta_{\tw}$ & $p$ & $\alpha_{\on}$ & $\alpha_{\tw}$ \\
\hline
1.1 & $0$ &  & $0$ & $({0},{1}]$ &  &  & $({0},{1}]$ &  &  &  &  &  & $({0},{\infty})$ &  &  \\
1.2 & $0$ &  &  & $({0},{1}]$ & $0$ & $0$ & $({0},{1}]$ &  &  &  &  &  & $({0},{\infty})$ &  &  \\
1.3 &  &  &  & $({0},{1}]$ & $0$ & $0$ & $({0},{1}]$ & $0$ & $0$ &  &  &  & $({0},{\infty})$ &  &  \\
1.4 & $0$ &  &  & $({0},{1}]$ & $0$ &  & $({0},{1}]$ &  &  & $-p - u$ &  &  & $({0},{\infty})$ &  &  \\
1.5 &  &  & $-b_{\on}$ & $({0},{1}]$ & $0$ &  & $({0},{1}]$ & $0$ & $0$ &  &  &  & $({0},{\infty})$ &  &  \\
1.6 &  &  & $-b_{\on}$ & $({0},{1}]$ &  &  & $({0},{1}]$ & $0$ & $0$ & $0$ &  &  & $[1,\infty)$ &  &  \\
1.7 &  &  & $-b_{\on}$ & $({0},{1}]$ & $0$ &  & $({0},{1}]$ & $0$ &  & $-p - v$ &  &  & $({0},{\infty})$ &  &  \\
1.8 &  &  & $-b_{\on}$ & $({0},{1}]$ & $0$ &  & $({0},{1}]$ &  & $0$ & $1 - p - v$ &  & $-\dfrac{2\beta_{\on}}{r_{\h}}$ & $({0},{\infty})$ &  & $0$ \\
1.9 &  &  &  & $({0},{1}]$ & $0$ &  & $({0},{1}]$ & $0$ & $0$ & $-p - u$ &  &  & $({0},{\infty})$ &  &  \\
1.10 &  &  &  & $({0},{1}]$ & $0$ & $0$ & $({0},{1}]$ & $0$ &  & $-p - v$ &  &  & $({0},{\infty})$ &  &  \\
1.11 &  &  &  & $({0},{1}]$ & $0$ &  & $u$ & $0$ &  & $-p - u$ &  &  & $({0},{\infty})$ &  &  \\
1.12 &  &  &  & $({0},{1}]$ & $0$ & $0$ & $({0},{1}]$ &  & $0$ & $1 - p - v$ &  & $-\dfrac{2\beta_{\on}}{r_{\h}}$ & $({0},{\infty})$ &  & $0$ \\[8pt]
\hline
\end{tabular}
\caption{Parameter values that satisfy the AFE for the case \( u > 0 \) and \( v > 0 \).}
\label{AFEcase1}
\end{table}

\subsection{$\boldsymbol{u > 0 \,\, \text{and} \,\, v < 0}$}

To ensure that we capture contributions from the dominant terms, we consider the case \( 0 < u \leq 1 \) and \( v \leq -1 \). In this regime, the leading-order terms are determined by the exponents \( -2 + v \), \( -2 + u \), and \( -p \), and we identify the branches that satisfy the AFE in Table~\ref{AFEcase2}.

\begin{table}[!h]
\centering
\begin{tabular}{|c|c|c|c|c|c|c|c|c|c|c|c|c|c|c|c|}
\hline
\multicolumn{4}{|c|}{\text{NGR}} & \multicolumn{3}{c|}{$\chi$} & \multicolumn{3}{c|}{$\psi$} & \multicolumn{3}{c|}{$A_1$}  & \multicolumn{3}{c|}{$A_2$} \\
\cline{1-16}
\# & $b_{\on}$ & $b_{\tw}$ & $b_{\tr}$ & $u$ & $\chi_{\n}$ & $\gamma_{\on}$ & $v$ & $\psi_{\n}$ & $\gamma_{\tw}$ & $q$ & $\beta_{\on}$ & $\beta_{\tw}$ & $p$ & $\alpha_{\on}$ & $\alpha_{\tw}$ \\
\hline
2.1 & 0 &   & 0         & $(0,1]$ &   &   & $(-\infty,-1]$ &   &   &   &   &   & $(0,\infty)$ & & \\
2.2 & 0 &   &           & $(0,1]$ & 0 & 0 & $(-\infty,-1]$ &   &   &   &   &   & $(0,\infty)$ & & \\
2.3 &   &   &           & $(0,1]$ & 0 & 0 & $(-\infty,-1]$ & 0 & 0 &   &   &   & $(0,\infty)$ & & \\
2.4 & 0 &   &           & $(0,1]$ & 0 &   & $(-\infty,-1]$ &   &   & $-p - u$ &   &   & $(0,\infty)$ & & \\
2.5 &   &   & $-b_{\on}$ & $(0,1]$ & 0 &   & $(-\infty,-1]$ & 0 & 0 &   &   &   & $(0,\infty)$ & & \\
2.6 &   &   & $-b_{\on}$ & $(0,1]$ &   &   & $(-\infty,-1]$ & 0 & 0 & 0 &   &   & $[1,\infty)$ & & \\
2.7 &   &   &           & $(0,1]$ & 0 &   & $(-\infty,-1]$ & 0 & 0 & $-p - u$ &   &   & $(0,\infty)$ & & \\
\hline
\end{tabular}
\caption{Parameter values that satisfy the AFE for the case \( u > 0 \) and \( v < 0 \). }
\label{AFEcase2}
\end{table}
\subsection{$\boldsymbol{u > 0 \,\, \text{and} \,\, v = 0}$}

To capture contributions from the dominant terms, we focus on the range \( 0 < u \leq 1 \). The leading-order terms are governed by the exponents \( -1 \), \( -p \), and \( -2 + p \). We identify the branches that satisfy the AFE as shown in Table~\ref{AFEcase3}.

\begin{table}[H]
\centering
\begin{tabular}{|c|c|c|c|c|c|c|c|c|c|c|c|c|c|c|c|}
\hline
\multicolumn{4}{|c|}{\text{NGR}} & \multicolumn{3}{c|}{$\chi$} & \multicolumn{3}{c|}{$\psi$} & \multicolumn{3}{c|}{$A_1$}  & \multicolumn{3}{c|}{$A_2$} \\
\cline{1-16}
\# & $b_{\on}$ & $b_{\tw}$ & $b_{\tr}$ & $u$ & $\chi_{\n}$ & $\gamma_{\on}$ & $v$ & $\psi_{\n}$ & $\gamma_{\tw}$ & $q$ & $\beta_{\on}$ & $\beta_{\tw}$ & $p$ & $\alpha_{\on}$ & $\alpha_{\tw}$ \\
\hline
3.1 & 0 &   & 0         & $(0,1]$ &   &   & 0 &   &   &   &   &   & $(0,\infty)$ & & \\
3.2 & 0 &   &           & $(0,1]$ & 0 & 0 & 0 &   &   &   &   &   & $(0,\infty)$ & & \\
3.3 &   &   &           & $(0,1]$ & 0 & 0 & 0 & 0 & 0 &   &   &   & $(0,\infty)$ & & \\
3.4 & 0 &   &           & $(0,1]$ & 0 &   & 0 &   &   & $-p - u$ &   &   & $(0,\infty)$ & & \\
3.5 &   &   & $-b_{\on}$ & $(0,1]$ & 0 &   & 0 & 0 & 0 &   &   &   & $(0,\infty)$ & & \\
3.6 &   &   & $-b_{\on}$ & $(0,1]$ &   &   & 0 & 0 & 0 & 0 &   &   & $[1,\infty)$ & & \\
3.7 &   &   & 0         & $(0,1]$ &   & 0 & 0 & 0 & 0 & $1 - p - u$ &   & $-\dfrac{2\beta_{\on}}{r_{\h}}$ & $(0,\infty)$ & & 0 \\
3.8 &   &   &           & $(0,1]$ & 0 &   & 0 & 0 & 0 & $-p - u$ &   &   & $(0,\infty)$ & & \\
\hline
\end{tabular}
\caption{Parameter values that satisfy the AFE for the case \( u > 0 \) and \( v = 0 \).}
\label{AFEcase3}
\end{table}
\subsection{$\boldsymbol{u < 0 \,\, \text{and} \,\, v > 0}$}

To include contributions from the dominant terms, we consider the range \( u \leq -1 \) and \( 0 < v \leq 1 \). The leading-order terms are governed by the exponents \( -2 + v \), \( 1 - p + u \), \( -2 + u \), and \( -1 - p \). We identify the branches that satisfy the AFE in Table~\ref{AFEcase4}.

\begin{table}[!h]
\centering
\begin{tabular}{|c|c|c|c|c|c|c|c|c|c|c|c|c|c|c|c|}
\hline
\multicolumn{4}{|c|}{\text{NGR}} & \multicolumn{3}{c|}{$\chi$} & \multicolumn{3}{c|}{$\psi$} & \multicolumn{3}{c|}{$A_1$}  & \multicolumn{3}{c|}{$A_2$} \\
\cline{1-16}
\# & $b_{\on}$ & $b_{\tw}$ & $b_{\tr}$ & $u$ & $\chi_{\n}$ & $\gamma_{\on}$ & $v$ & $\psi_{\n}$ & $\gamma_{\tw}$ & $q$ & $\beta_{\on}$ & $\beta_{\tw}$ & $p$ & $\alpha_{\on}$ & $\alpha_{\tw}$ \\
\hline
4.1 & 0 &   & 0         & $(-\infty,-1]$ &   &   & $(0,1]$ &   &   &   &   &   & $(0,\infty)$ & & \\
4.2 & 0 &   &           & $(-\infty,-1]$ & 0 & 0 & $(0,1]$ &   &   &   &   &   & $(0,\infty)$ & & \\
4.3 &   &   &           & $(-\infty,-1]$ & 0 & 0 & $(0,1]$ & 0 & 0 &   &   &   & $(0,\infty)$ & & \\
4.4 &   &   &           & $(-\infty,-1]$ & 0 & 0 & $(0,1]$ & 0 &   & $-p - v$ &   &   & $(0,\infty)$ & & \\
\hline
\end{tabular}
\caption{Parameter values that satisfy the AFE for the case $u < 0$ and $v > 0$.}
\label{AFEcase4}
\end{table}
\subsection{$\boldsymbol{u < 0 \,\, \text{and} \,\, v < 0}$}

In order to include contributions from the dominant terms, we restrict our attention to the range \( u \leq -1 \) and \( v \leq -1 \). The leading-order terms are determined by the exponents \( -2 + v \), \( -1 - p + u \), \( -2 + u \), and \( -1 - p + v \). In the search for branches that satisfy the AFE, we identify the cases in Table~\ref{AFEcase5}.

\begin{table}[!h]
\centering
\begin{tabular}{|c|c|c|c|c|c|c|c|c|c|c|c|c|c|c|c|}
\hline
\multicolumn{4}{|c|}{\text{NGR}} & \multicolumn{3}{c|}{$\chi$} & \multicolumn{3}{c|}{$\psi$} & \multicolumn{3}{c|}{$A_1$}  & \multicolumn{3}{c|}{$A_2$} \\
\cline{1-16}
\# & $b_{\on}$ & $b_{\tw}$ & $b_{\tr}$ & $u$ & $\chi_{\n}$ & $\gamma_{\on}$ & $v$ & $\psi_{\n}$ & $\gamma_{\tw}$ & $q$ & $\beta_{\on}$ & $\beta_{\tw}$ & $p$ & $\alpha_{\on}$ & $\alpha_{\tw}$ \\
\hline
5.1 & 0 &   & 0         & $(-\infty,-1]$ &   &   & $(-\infty,-1]$ &   &   &   &   &   & $(0,\infty)$ & & \\
5.2 & 0 &   &           & $(-\infty,-1]$ & 0 & 0 & $(-\infty,-1]$ &   &   &   &   &   & $(0,\infty)$ & & \\
5.3 &   &   &           & $(-\infty,-1]$ & 0 & 0 & $(-\infty,-1]$ & 0 & 0 &   &   &   & $(0,\infty)$ & & \\
\hline
\end{tabular}
\caption{Parameter values that satisfy the AFE for the case \( u < 0 \) and \( v < 0 \). }
\label{AFEcase5}
\end{table}
\subsection{$\boldsymbol{u < 0 \,\, \text{and} \,\, v = 0}$}

In order to include contributions from the dominant terms, we restrict our attention to the range \( u \leq -1 \). The leading-order terms are determined by the exponents \( -1 - p + u \), \( -2 + u \), \( -2p \), and \( -1 - p \). In the search for branches that satisfy the AFE, we identify the cases in  Table~\ref{AFEcase6}.

\begin{table}[!h]
\centering
\begin{tabular}{|c|c|c|c|c|c|c|c|c|c|c|c|c|c|c|c|}
\hline
\multicolumn{4}{|c|}{\text{NGR}} & \multicolumn{3}{c|}{$\chi$} & \multicolumn{3}{c|}{$\psi$} & \multicolumn{3}{c|}{$A_1$}  & \multicolumn{3}{c|}{$A_2$} \\
\cline{1-16}
\# & $b_{\on}$ & $b_{\tw}$ & $b_{\tr}$ & $u$ & $\chi_{\n}$ & $\gamma_{\on}$ & $v$ & $\psi_{\n}$ & $\gamma_{\tw}$ & $q$ & $\beta_{\on}$ & $\beta_{\tw}$ & $p$ & $\alpha_{\on}$ & $\alpha_{\tw}$ \\
\hline
6.1 & 0 &   & 0         & $(-\infty,-1]$ &   &   & 0 &   &   &   &   &   & $(0,\infty)$ & & \\
6.2 & 0 &   &           & $(-\infty,-1]$ & 0 & 0 & 0 &   &   &   &   &   & $(0,\infty)$ & & \\
6.3 &   &   &           & $(-\infty,-1]$ & 0 & 0 & 0 & 0 & 0 &   &   &   & $(0,\infty)$ & & \\
6.4 &   &   &           & $(-\infty,-1]$ & 0 & 0 & 0 & 0 &   & $-p$ &   & $a \beta_{\on}$ & $(0,1)$ & & $-\left(\dfrac{2}{r_{\h}} + a\right)\alpha_{\on}$ \\[8pt]
\hline
\end{tabular}
\caption{Parameter values that satisfy the AFE for the case \( u < 0 \) and \( v = 0 \), where $a=\beta_{\tw}/\beta_{\on}$ is an arbitrary constant. }
\label{AFEcase6}
\end{table}

\subsection{$\boldsymbol{u = 0 \,\, \text{and} \,\, v > 0}$}

In order to include contributions from the dominant terms, we restrict our attention to the range \( 0 < v \leq 1 \). The leading-order terms are determined by the exponents \( -2 + v \), \( -1 \), and \( -1 - p \). In the search for branches that satisfy the AFE, we identify the cases in Table~\ref{AFEcase7}.

\begin{table}[!h]
\centering
\begin{tabular}{|c|c|c|c|c|c|c|c|c|c|c|c|c|c|c|c|}
\hline
\multicolumn{4}{|c|}{\text{NGR}} & \multicolumn{3}{c|}{$\chi$} & \multicolumn{3}{c|}{$\psi$} & \multicolumn{3}{c|}{$A_1$}  & \multicolumn{3}{c|}{$A_2$} \\
\cline{1-16}
\# & $b_{\on}$ & $b_{\tw}$ & $b_{\tr}$ & $u$ & $\chi_{\n}$ & $\gamma_{\on}$ & $v$ & $\psi_{\n}$ & $\gamma_{\tw}$ & $q$ & $\beta_{\on}$ & $\beta_{\tw}$ & $p$ & $\alpha_{\on}$ & $\alpha_{\tw}$ \\
\hline
7.1 & 0 &   & 0         & 0 &   &   & $(0,1]$ &   &   &   &   &   & $(0,\infty)$ & & \\
7.2 & 0 &   &           & 0 & $n\pi$ & 0 & $(0,1]$ &   &   &   &   &   & $(0,\infty)$ & & \\
7.3 &   &   &           & 0 & $n\pi$ & 0 & $(0,1]$ & 0 & 0 &   &   &   & $(0,\infty)$ & & \\
7.4 &   &   & $-b_{\on}$ & 0 & $n\pi$ &   & $(0,1]$ & 0 & 0 & 0 &   &   & $(0,\tfrac{1}{2})$ & & \\
7.5 &   &   & 0         & 0 & $n\pi$ &   & $(0,1]$ & 0 & 0 & $-p$ &   & $a\beta_{\on}$ & $(0,\infty)$ & & $-\left(\dfrac{2}{r_{\h}} + a\right)\alpha_{\on}$ \\
7.6 &   &   &           & 0 & $n\pi$ & 0 & $(0,1]$ & 0 &   & $-p - v$ &   &   & $(0,\infty)$ & & \\
\hline
\end{tabular}
\caption{Parameter values that satisfy the AFE for the case \( u = 0 \) and \( v > 0 \), where $a=\beta_{\tw}/\beta_{\on}$ is an arbitrary constant.}
\label{AFEcase7}
\end{table}

\subsection{$\boldsymbol{u = 0 \,\, \text{and} \,\, v < 0}$}

In order to include contributions from the dominant terms, we restrict our attention to the range \( v \leq -1 \). The leading-order terms are determined by the exponents \( -2 + v \), \( -1 - p + v \), and \( -2p \). In the search for branches that satisfy the AFE, we identify the cases in Table~\ref{AFEcase8}.

\begin{table}[!h]
\centering
\begin{tabular}{|c|c|c|c|c|c|c|c|c|c|c|c|c|c|c|c|}
\hline
\multicolumn{4}{|c|}{\text{NGR}} & \multicolumn{3}{c|}{$\chi$} & \multicolumn{3}{c|}{$\psi$} & \multicolumn{3}{c|}{$A_1$}  & \multicolumn{3}{c|}{$A_2$} \\
\cline{1-16}
\# & $b_{\on}$ & $b_{\tw}$ & $b_{\tr}$ & $u$ & $\chi_{\n}$ & $\gamma_{\on}$ & $v$ & $\psi_{\n}$ & $\gamma_{\tw}$ & $q$ & $\beta_{\on}$ & $\beta_{\tw}$ & $p$ & $\alpha_{\on}$ & $\alpha_{\tw}$ \\
\hline
8.1 & 0 &   & 0         & 0 &   &   & $(-\infty,-1]$ &   &   &   &   &   & $(0,\infty)$ & & \\
8.2 & 0 &   &           & 0 & $n\pi$ & 0 & $(-\infty,-1]$ &   &   &   &   &   & $(0,\infty)$ & & \\
8.3 &   &   &           & 0 & $n\pi$ & 0 & $(-\infty,-1]$ & 0 & 0 &   &   &   & $(0,\infty)$ & & \\
8.4 &   &   & $-b_{\on}$ & 0 & $n\pi$ &   & $(-\infty,-1]$ & 0 & 0 & 0 &   &   & $(0,\tfrac{1}{2})$ & & \\
8.5 &   &   & 0         & 0 & $n\pi$ &   & $(-\infty,-1]$ & 0 & 0 & $-p$ &   & $a\beta_{\on}$ & $(0,1)$ & & $-\left(\dfrac{2}{r_{\h}} + a\right)\alpha_{\on}$ \\[8pt]
\hline
\end{tabular}
\caption{Parameter values that satisfy the AFE for the case \( u = 0 \) and \( v < 0 \), where $a=\beta_{\tw}/\beta_{\on}$ is an arbitrary constant.}
\label{AFEcase8}
\end{table}

\section{Results of the SFE perturbative analysis}\label{SFEp}

In this section, we summarize the results of the perturbative analysis of the SFE. For a complete and detailed treatment, refer to \citep{thesis}. Note that all cases are presented here except for the case $u=v=0$, which was outlined in Section~\ref{SFEp9}.
\subsection{$\boldsymbol{0<u\leq 1 \,\, \text{and} \,\, 0<v\leq 1}$}

The leading-order terms are determined by the exponents \( -2 \), \( -2p \), and \( -1 - p \). Considering the cases, $0<p<1$, \( p = 1 \), and $p>1$, the requirement that their corresponding coefficients vanish guides our analysis. We identify the cases in Table~\ref{FEcase1} that satisfy both the AFE and the SFE.

\begin{table}[!h]
\centering
\begin{tabular}{|c|c|c|c|c|c|c|c|c|c|c|c|c|c|}
\hline
\multicolumn{4}{|c|}{\text{NGR}} & \multicolumn{2}{c|}{$\chi$} & \multicolumn{2}{c|}{$\psi$} & \multicolumn{3}{c|}{$A_1$} & \multicolumn{3}{c|}{$A_2$} \\
\cline{1-14}
\# & $b_{\on}$ & $b_{\tw}$ & $b_{\tr}$ & $\chi_{\n}$ & $\gamma_{\on}$ & $\psi_{\n}$ & $\gamma_{\tw}$ & $q$ & $\beta_{\on}$ & $\beta_{\tw}$ & $p$ & $\alpha_{\on}$ & $\alpha_{\tw}$ \\
\hline
1.1.1 & 0 &   & 0 &  &  &  &  & $\frac{1}{2}$ &  & $\bar{\beta}_{\on}$ & $\frac{1}{2}$ & $\dfrac{\delta}{\sqrt{r_{\h}}}$ & \\[11pt]
1.2.1 & 0 &   &  & $0$ & $0$ &  &  & $\frac{1}{2}$ &  & $\bar{\beta}_{\on}$ & $\frac{1}{2}$ & $\dfrac{\delta}{\sqrt{r_{\h}}}$ & \\[11pt]
1.3.1 &   & $0$ &  & $0$ & $0$ & $0$ & $0$ & $\dfrac{-2}{\alpha_{\on} r_{\h}}$ &  & $\dfrac{-2\beta_{\on}}{r_{\h}}$ & $1$ &  & $\dfrac{-\alpha_{\on}}{r_{\h}}$ \\[11pt]
1.3.2 &   & $3 b_{\on}$ &  & $0$ & $0$ & $0$ & $0$ & $\dfrac{1}{\alpha_{\on} r_{\h}}$ &  & $\dfrac{\beta_{\on}}{r_{\h}}$ & $1$ &  & $\dfrac{-\alpha_{\on}}{r_{\h}}$ \\[11pt]
1.3.3 &   & $4 b_{\on}$ &  & $0$ & $0$ & $0$ & $0$ & $0$ &  & $0$ & $1$ & $\dfrac{-1}{r_{\h}}$ & \\[11pt]
1.3.4 &   & $4 b_{\on}$ &  & $0$ & $0$ & $0$ & $0$ & $0$ &  & $\dfrac{4\beta_{\on}}{r_{\h}}$ & $1$ & $\dfrac{1}{r_{\h}}$ & \\[11pt]
1.9.1 &   & $3 b_{\on}$ &  & $0$ &  & $0$ & $0$ & $-1 - u$ &  & $\dfrac{\beta_{\on}}{r_{\h}}$ & $1$ & $\dfrac{-r_{\h}^{-1}}{1 + u}$ & $\dfrac{r_{\h}^{-2}}{1 + u}$ \\[11pt]
1.10.1 &   & $3 b_{\on}$ &  & $0$ & $0$ & $0$ &  & $-1 - v$ &  & $\dfrac{\beta_{\on}}{r_{\h}}$ & $1$ & $\dfrac{-r_{\h}^{-1}}{1 + v}$ & $\dfrac{r_{\h}^{-2}}{1 + v}$ \\[11pt]
\hline
\end{tabular}
\caption{Parameter values that satisfy both the AFE and SFE for the case \( 0 < u \leq 1 \) and \( 0 < v \leq 1 \), where $\delta=\pm 1$.}
\label{FEcase1}
\end{table}

\subsection{$\boldsymbol{0<u\leq 1 \,\, \text{and} \,\, v\leq -1}$}

The leading-order terms are determined by the exponents \( -2 + 2v \) and \( -2p \). We examine different configurations, and the requirement that their corresponding coefficients vanish guides our analysis. In the search for branches that satisfy both the AFE and the SFE, we identify the cases in Table~\ref{FEcase2}.

\begin{table}[!h]
\centering
\begin{tabular}{|c|c|c|c|c|c|c|c|c|c|c|c|c|c|}
\hline
\multicolumn{4}{|c|}{\text{NGR}} & \multicolumn{2}{c|}{$\chi$} & \multicolumn{2}{c|}{$\psi$} & \multicolumn{3}{c|}{$A_1$} & \multicolumn{3}{c|}{$A_2$} \\
\cline{1-14}
\# & $b_{\on}$ & $b_{\tw}$ & $b_{\tr}$ & $\chi_{\n}$ & $\gamma_{\on}$ & $\psi_{\n}$ & $\gamma_{\tw}$ & $q$ & $\beta_{\on}$ & $\beta_{\tw}$ & $p$ & $\alpha_{\on}$ & $\alpha_{\tw}$ \\
\hline
2.1.1 & $0$ &  & $0$ &  &  &  &  & $\frac{1}{2}$ &  & $\bar{\beta}_{\on}$ & $\frac{1}{2}$ & $\dfrac{\delta}{\sqrt{r_{\h}}}$ & \\[11pt]
2.2.1 & $0$ &  &  & $0$ & $0$ &  &  & $\frac{1}{2}$ &  & $\bar{\beta}_{\on}$ & $\frac{1}{2}$ & $\dfrac{\delta}{\sqrt{r_{\h}}}$ & \\[11pt]
2.3.1 &  & $0$ &  & $0$ & $0$ & $0$ & $0$ & $-\dfrac{2}{\alpha_{\on} r_{\h}}$ &  & $-\dfrac{2\beta_{\on}}{r_{\h}}$ & $1$ &  & $-\dfrac{\alpha_{\on}}{r_{\h}}$ \\[11pt]
2.3.2 &  & $3 b_{\on}$ &  & $0$ & $0$ & $0$ & $0$ & $\dfrac{1}{\alpha_{\on} r_{\h}}$ &  & $\dfrac{\beta_{\on}}{r_{\h}}$ & $1$ &  & $-\dfrac{\alpha_{\on}}{r_{\h}}$ \\[11pt]
2.3.3 &  & $4 b_{\on}$ &  & $0$ & $0$ & $0$ & $0$ & $0$ &  & $0$ & $1$ & $-\dfrac{1}{r_{\h}}$ & \\[11pt]
2.3.4 &  & $4 b_{\on}$ &  & $0$ & $0$ & $0$ & $0$ & $0$ &  & $\dfrac{4\beta_{\on}}{r_{\h}}$ & $1$ & $\dfrac{1}{r_{\h}}$ &    \\[11pt]
2.7.1 &  & $3 b_{\on}$ &  & $0$ &  & $0$ & $0$ & $-1 - u$ &  & $\dfrac{\beta_{\on}}{r_{\h}}$ & $1$ & $\dfrac{-r_{\h}^{-1}}{1 + u}$ & $\dfrac{r_{\h}^{-2}}{1 + u}$ \\[11pt]
\hline
\end{tabular}
\caption{Parameter values that satisfy both the AFE and SFE for the case \( 0 < u \leq 1 \) and \( v\leq -1 \), where $\delta=\pm 1$.}
\label{FEcase2}
\end{table}

\subsection{$\boldsymbol{0<u\leq 1 \,\, \text{and} \,\, v=0}$}

The leading-order terms are determined by the exponents \( -2 \), \( -2p \), and \( -1 - p \). Considering the cases \( 0 < p < 1 \), \( p = 1 \), and \( p > 1 \), the requirement that their corresponding coefficients vanish guides our analysis. In the search for branches that satisfy both the AFE and the SFE, we identify the cases in Table~\ref{FEcase3}.

\begin{table}[!h]
\centering
\begin{tabular}{|c|c|c|c|c|c|c|c|c|c|c|c|c|c|}
\hline
\multicolumn{4}{|c|}{\text{NGR}} & \multicolumn{2}{c|}{$\chi$} & \multicolumn{2}{c|}{$\psi$} & \multicolumn{3}{c|}{$A_1$} & \multicolumn{3}{c|}{$A_2$} \\
\cline{1-14}
\# & $b_{\on}$ & $b_{\tw}$ & $b_{\tr}$ & $\chi_{\n}$ & $\gamma_{\on}$ & $\psi_{\n}$ & $\gamma_{\tw}$ & $q$ & $\beta_{\on}$ & $\beta_{\tw}$ & $p$ & $\alpha_{\on}$ & $\alpha_{\tw}$ \\
\hline
3.1.1 & $0$ &  & $0$ &  &  &  &  & $\frac{1}{2}$ &  & $\bar{\beta}_{\on}$ & $\frac{1}{2}$ & $\dfrac{\delta}{\sqrt{r_{\h}}}$ & \\[11pt]
3.2.1 & $0$ &  &  & $0$ & $0$ &  &  & $\frac{1}{2}$ &  & $\bar{\beta}_{\on}$ & $\frac{1}{2}$ & $\dfrac{\delta}{\sqrt{r_{\h}}}$ & \\[11pt]
3.3.1 &  & $0$ &  & $0$ & $0$ & $0$ & $0$ & $-\dfrac{2}{\alpha_{\on} r_{\h}}$ &  & $-\dfrac{2\beta_{\on}}{r_{\h}}$ & $1$ &  & $-\dfrac{\alpha_{\on}}{r_{\h}}$ \\[11pt]
3.3.2 &  & $3 b_{\on}$ &  & $0$ & $0$ & $0$ & $0$ & $\dfrac{1}{\alpha_{\on} r_{\h}}$ &  & $\dfrac{\beta_{\on}}{r_{\h}}$ & $1$ &  & $-\dfrac{\alpha_{\on}}{r_{\h}}$ \\[11pt]
3.3.3 &  & $4 b_{\on}$ &  & $0$ & $0$ & $0$ & $0$ & $0$ &  & $0$ & $1$ & $-\dfrac{1}{r_{\h}}$ & \\[11pt]
3.3.4 &  & $4 b_{\on}$ &  & $0$ & $0$ & $0$ & $0$ & $0$ &  & $\dfrac{4\beta_{\on}}{r_{\h}}$ & $1$ & $\dfrac{1}{r_{\h}}$ & \\[11pt]
3.8.1 &  & $3 b_{\on}$ &  & $0$ &  & $0$ & $0$ & $-1 - u$ &  & $\dfrac{\beta_{\on}}{r_{\h}}$ & $1$ & $\dfrac{-r_{\h}^{-1}}{1 + u}$ & $\dfrac{r_{\h}^{-2}}{1 + u}$ \\[11pt]
\hline
\end{tabular}
\caption{Parameter values that satisfy both the AFE and SFE for the case \( 0 < u \leq 1 \) and \( v=0 \). }
\label{FEcase3}
\end{table}

\subsection{$\boldsymbol{u \leq -1 \,\, \text{and} \,\, 0< v \leq 1 }$}

The leading-order terms are determined by the exponents \( -2 + 2u \), \( -2p \), and \( -1 - p + u \). Our analysis is guided by considering all possible configurations of these exponents and requiring that their corresponding coefficients vanish. In the search for branches that satisfy both the AFE and the SFE, we identify the cases in Table~\ref{FEcase4}.

\begin{table}[!h]
\centering
\begin{tabular}{|c|c|c|c|c|c|c|c|c|c|c|c|c|c|}
\hline
\multicolumn{4}{|c|}{\text{NGR}} & \multicolumn{2}{c|}{$\chi$} & \multicolumn{2}{c|}{$\psi$} & \multicolumn{3}{c|}{$A_1$} & \multicolumn{3}{c|}{$A_2$} \\
\cline{1-14}
\# & $b_{\on}$ & $b_{\tw}$ & $b_{\tr}$ & $\chi_{\n}$ & $\gamma_{\on}$ & $\psi_{\n}$ & $\gamma_{\tw}$ & $q$ & $\beta_{\on}$ & $\beta_{\tw}$ & $p$ & $\alpha_{\on}$ & $\alpha_{\tw}$ \\
\hline
4.1.1 & $0$ &  & $0$ &  &  &  &  & $\frac{1}{2}$ &  & $\bar{\beta}_{\on}$ & $\frac{1}{2}$ & $\dfrac{\delta}{\sqrt{r_{\h}}}$ & \\[11pt]
4.2.1 & $0$ &  &  & $0$ & $0$ &  &  & $\frac{1}{2}$ &  & $\bar{\beta}_{\on}$ & $\frac{1}{2}$ & $\dfrac{\delta}{\sqrt{r_{\h}}}$ & \\[11pt]
4.3.1 &  & $0$ &  & $0$ & $0$ & $0$ & $0$ & $-\dfrac{2}{\alpha_{\on} r_{\h}}$ &  & $-\dfrac{2\beta_{\on}}{r_{\h}}$ & $1$ &  & $-\dfrac{\alpha_{\on}}{r_{\h}}$ \\[11pt]
4.3.2 &  & $3 b_{\on}$ &  & $0$ & $0$ & $0$ & $0$ & $\dfrac{1}{\alpha_{\on} r_{\h}}$ &  & $\dfrac{\beta_{\on}}{r_{\h}}$ & $1$ &  & $-\dfrac{\alpha_{\on}}{r_{\h}}$ \\[11pt]
4.3.3 &  & $4 b_{\on}$ &  & $0$ & $0$ & $0$ & $0$ & $0$ &  & $0$ & $1$ & $-\dfrac{1}{r_{\h}}$ & \\[11pt]
4.3.4 &  & $4 b_{\on}$ &  & $0$ & $0$ & $0$ & $0$ & $0$ &  & $\dfrac{4\beta_{\on}}{r_{\h}}$ & $1$ & $\dfrac{1}{r_{\h}}$ & \\[11pt]
4.4.1 &  & $3 b_{\on}$ &  & $0$ & $0$ & $0$ &  & $-1 - v$ &  & $\dfrac{\beta_{\on}}{r_{\h}}$ & $1$ & $\dfrac{-r_{\h}^{-1}}{1 + v}$ & $\dfrac{r_{\h}^{-2}}{1 + v}$ \\[11pt]
\hline
\end{tabular}
\caption{Parameter values that satisfy both the AFE and SFE for the case \( u \leq -1 \) and \( 0<v\leq 1 \), where $\delta=\pm 1$.}
\label{FEcase4}
\end{table}

\subsection{$\boldsymbol{u \leq -1 \,\, \text{and} \,\, v \leq -1}$}

The leading-order terms are determined by the exponents \( -2 + 2u \), \( -2 + 2v \), \( -2p \), \( -1 - p + u \), and \( -1 - p + v \). Our analysis is guided by considering all possible configurations of these exponents and requiring that their corresponding coefficients vanish. In the search for branches that satisfy both the AFE and the SFE, we identify the cases in Table~\ref{FEcase5}.

\begin{table}[!h]
\centering
\begin{tabular}{|c|c|c|c|c|c|c|c|c|c|c|c|c|c|}
\hline
\multicolumn{4}{|c|}{\text{NGR}} & \multicolumn{2}{c|}{$\chi$} & \multicolumn{2}{c|}{$\psi$} & \multicolumn{3}{c|}{$A_1$} & \multicolumn{3}{c|}{$A_2$} \\
\cline{1-14}
\# & $b_{\on}$ & $b_{\tw}$ & $b_{\tr}$ & $\chi_{\n}$ & $\gamma_{\on}$ & $\psi_{\n}$ & $\gamma_{\tw}$ & $q$ & $\beta_{\on}$ & $\beta_{\tw}$ & $p$ & $\alpha_{\on}$ & $\alpha_{\tw}$ \\
\hline
5.1.1 & $0$ &  & $0$ &  &  &  &  & $\frac{1}{2}$ &  & $\bar{\beta}_{\on}$ & $\frac{1}{2}$ & $\dfrac{\delta}{\sqrt{r_{\h}}}$ & \\[11pt]
5.2.1 & $0$ &  &  & $0$ & $0$ &  &  & $\frac{1}{2}$ &  & $\bar{\beta}_{\on}$ & $\frac{1}{2}$ & $\dfrac{\delta}{\sqrt{r_{\h}}}$ & \\[11pt]
5.3.1 &  & $0$ &  & $0$ & $0$ & $0$ & $0$ & $-\frac{2}{\alpha_{\on} r_{\h}}$ &  & $-\dfrac{2\beta_{\on}}{r_{\h}}$ & $1$ &  & $-\dfrac{\alpha_{\on}}{r_{\h}}$ \\[11pt]
5.3.2 &  & $3 b_{\on}$ &  & $0$ & $0$ & $0$ & $0$ & $\frac{1}{\alpha_{\on} r_{\h}}$ &  & $\dfrac{\beta_{\on}}{r_{\h}}$ & $1$ &  & $-\dfrac{\alpha_{\on}}{r_{\h}}$ \\[11pt]
5.3.3 &  & $4 b_{\on}$ &  & $0$ & $0$ & $0$ & $0$ & $0$ &  & $0$ & $1$ & $-\dfrac{1}{r_{\h}}$ & \\[11pt]
5.3.4 &  & $4 b_{\on}$ &  & $0$ & $0$ & $0$ & $0$ & $0$ &  & $\dfrac{4\beta_{\on}}{r_{\h}}$ & $1$ & $\dfrac{1}{r_{\h}}$ & \\[11pt]
\hline
\end{tabular}
\caption{Parameter values that satisfy both the AFE and SFE for the case \( u \leq -1 \) and \( v\leq -1 \), where $\delta=\pm 1$. }
\label{FEcase5}
\end{table}

\subsection{$\boldsymbol{u \leq -1 \,\, \text{and} \,\, v = 0}$}

The leading-order terms are determined by the exponents \( -2 + 2u \), \( -2p \), and \( -1 - p + u \). Our analysis is guided by considering all possible values of these exponents and requiring that their corresponding coefficients vanish. In the search for branches that satisfy both the AFE and the SFE, we identify the cases in Table~\ref{FEcase6}.

\begin{table}[!h]
\centering
\begin{tabular}{|c|c|c|c|c|c|c|c|c|c|c|c|c|c|}
\hline
\multicolumn{4}{|c|}{\text{NGR}} & \multicolumn{2}{c|}{$\chi$} & \multicolumn{2}{c|}{$\psi$} & \multicolumn{3}{c|}{$A_1$} & \multicolumn{3}{c|}{$A_2$} \\
\cline{1-14}
\# & $b_{\on}$ & $b_{\tw}$ & $b_{\tr}$ & $\chi_{\n}$ & $\gamma_{\on}$ & $\psi_{\n}$ & $\gamma_{\tw}$ & $q$ & $\beta_{\on}$ & $\beta_{\tw}$ & $p$ & $\alpha_{\on}$ & $\alpha_{\tw}$ \\
\hline
6.1.1 & $0$ &  & $0$ &  &  &  &  & $\frac{1}{2}$ &  & $\bar{\beta}_{\on}$ & $\frac{1}{2}$ & $\dfrac{\delta}{\sqrt{r_{\h}}}$ & \\[11pt]
6.2.1 & $0$ &  &  & $0$ & $0$ &  &  & $\frac{1}{2}$ &  & $\bar{\beta}_{\on}$ & $\frac{1}{2}$ & $\dfrac{\delta}{\sqrt{r_{\h}}}$ & \\[11pt]
6.3.1 &  & $0$ &  & $0$ & $0$ & $0$ & $0$ & $-\dfrac{2}{\alpha_{\on} r_{\h}}$ &  & $-\dfrac{2\beta_{\on}}{r_{\h}}$ & $1$ &  & $-\dfrac{\alpha_{\on}}{r_{\h}}$ \\[11pt]
6.3.2 &  & $3 b_{\on}$ &  & $0$ & $0$ & $0$ & $0$ & $\frac{1}{\alpha_{\on} r_{\h}}$ &  & $\dfrac{\beta_{\on}}{r_{\h}}$ & $1$ &  & $-\dfrac{\alpha_{\on}}{r_{\h}}$ \\[11pt]
6.3.3 &  & $4 b_{\on}$ &  & $0$ & $0$ & $0$ & $0$ & $0$ &  & $0$ & $1$ & $-\dfrac{1}{r_{\h}}$ & \\[11pt]
6.3.4 &  & $4 b_{\on}$ &  & $0$ & $0$ & $0$ & $0$ & $0$ &  & $\dfrac{4\beta_{\on}}{r_{\h}}$ & $1$ & $\dfrac{1}{r_{\h}}$ & \\[11pt]
\hline
\end{tabular}
\caption{Parameter values that satisfy both the AFE and SFE for the case \( u \leq -1 \) and \( v=0 \), where $\delta=\pm 1$.}
\label{FEcase6}
\end{table}

\subsection{$\boldsymbol{u = 0 \,\, \text{and} \,\, 0< v \leq 1}$}

The leading-order terms are determined by the exponents \( -2 \), \( -2p \), and \( -1 - p \). Our analysis is guided by considering the possible values of \( p \) and requiring that the corresponding coefficients vanish. In the search for branches that satisfy both the AFE and the SFE, we identify the cases in Table~\ref{FEcase7}.

\begin{table}[!h]
\centering
\begin{tabular}{|c|c|c|c|c|c|c|c|c|c|c|c|c|c|}
\hline
\multicolumn{4}{|c|}{\text{NGR}} & \multicolumn{2}{c|}{$\chi$} & \multicolumn{2}{c|}{$\psi$} & \multicolumn{3}{c|}{$A_1$} & \multicolumn{3}{c|}{$A_2$} \\
\cline{1-14}
\# & $b_{\on}$ & $b_{\tw}$ & $b_{\tr}$ & $\chi_{\n}$ & $\gamma_{\on}$ & $\psi_{\n}$ & $\gamma_{\tw}$ & $q$ & $\beta_{\on}$ & $\beta_{\tw}$ & $p$ & $\alpha_{\on}$ & $\alpha_{\tw}$ \\
\hline
7.1.1 & $0$ &  & $0$ &  &  &  &  & $\frac{1}{2}$ &  & $\bar{\beta}_{\on}$ & $\frac{1}{2}$ & $\dfrac{\delta}{\sqrt{r_{\h}}}$ & \\[11pt]
7.2.1 & $0$ &  &  & $n\pi$ & $0$ &  &  & $\frac{1}{2}$ &  & $\bar{\beta}_{\on}$ & $\frac{1}{2}$ & $\dfrac{\delta}{\sqrt{r_{\h}}}$ & \\[11pt]
7.3.1 &  & $0$ &  & $n\pi$ & $0$ & $0$ & $0$ & $-\dfrac{2\delta}{\alpha_{\on} r_{\h}}$ &  & $-\dfrac{2\beta_{\on}}{r_{\h}}$ & $1$ &  & $-\dfrac{\alpha_{\on}}{r_{\h}}$ \\[11pt]
7.3.2 &  & $3 b_{\on}$ &  & $n\pi$ & $0$ & $0$ & $0$ & $\dfrac{\delta}{\alpha_{\on} r_{\h}}$ &  & $\dfrac{\beta_{\on}}{r_{\h}}$ & $1$ &  & $-\dfrac{\alpha_{\on}}{r_{\h}}$ \\[11pt]
7.3.3 &  & $4 b_{\on}$ &  & $n\pi$ & $0$ & $0$ & $0$ & $0$ &  & $0$ & $1$ & $-\dfrac{\delta}{r_{\h}}$ & \\[11pt]
7.3.4 &  & $4 b_{\on}$ &  & $n\pi$ & $0$ & $0$ & $0$ & $0$ &  & $\dfrac{4\beta_{\on}}{r_{\h}}$ & $1$ & $\dfrac{\delta}{r_{\h}}$ & \\[11pt]
7.6.1 &  & $3 b_{\on}$ &  & $n\pi$ & $0$ & $0$ &  & $-1 - v$ &  & $\dfrac{\beta_{\on}}{r_{\h}}$ & $1$ & $\dfrac{-\delta r_{\h}^{-1}}{1 + v}$ & $\dfrac{\delta r_{\h}^{-2}}{1 + v}$ \\[11pt]
\hline
\end{tabular}
\caption{Parameter values that satisfy both the AFE and SFE for the case \( u =0 \) and \( 0<v\leq 1 \), where $\delta=\pm 1$.}
\label{FEcase7}
\end{table}

\subsection{$\boldsymbol{u = 0 \,\, \text{and} \,\, v \leq -1}$}

The leading-order terms are determined by the exponents \( -2 + 2v \), \( -2p \), and \( -1 - p + v \). Our analysis is guided by considering all possible values of these exponents and requiring that their corresponding coefficients vanish. In the search for branches that satisfy both the AFE and the SFE, we identify the cases in Table~\ref{FEcase8}.

\begin{table}[!h]
\centering
\begin{tabular}{|c|c|c|c|c|c|c|c|c|c|c|c|c|c|}
\hline
\multicolumn{4}{|c|}{\text{NGR}} & \multicolumn{2}{c|}{$\chi$} & \multicolumn{2}{c|}{$\psi$} & \multicolumn{3}{c|}{$A_1$} & \multicolumn{3}{c|}{$A_2$} \\
\cline{1-14}
\# & $b_{\on}$ & $b_{\tw}$ & $b_{\tr}$ & $\chi_{\n}$ & $\gamma_{\on}$ & $\psi_{\n}$ & $\gamma_{\tw}$ & $q$ & $\beta_{\on}$ & $\beta_{\tw}$ & $p$ & $\alpha_{\on}$ & $\alpha_{\tw}$ \\
\hline
8.1.1 & $0$ &  & $0$ &  &  &  &  & $\frac{1}{2}$ &  & $\bar{\beta}_{\on}$ & $\frac{1}{2}$ & $\dfrac{\delta}{\sqrt{r_{\h}}}$ & \\[11pt]
8.2.1 & $0$ &  &  & $n\pi$ & $0$ &  &  & $\frac{1}{2}$ &  & $\bar{\beta}_{\on}$ & $\frac{1}{2}$ & $\dfrac{\delta}{\sqrt{r_{\h}}}$ & \\[11pt]
8.3.1 &  & $0$ &  & $n\pi$ & $0$ & $0$ & $0$ & $-\dfrac{2\delta}{\alpha_{\on} r_{\h}}$ &  & $-\dfrac{2\beta_{\on}}{r_{\h}}$ & $1$ &  & $-\dfrac{\alpha_{\on}}{r_{\h}}$ \\[11pt]
8.3.2 &  & $3 b_{\on}$ &  & $n\pi$ & $0$ & $0$ & $0$ & $\dfrac{\delta}{\alpha_{\on} r_{\h}}$ &  & $\dfrac{\beta_{\on}}{r_{\h}}$ & $1$ &  & $-\dfrac{\alpha_{\on}}{r_{\h}}$ \\[11pt]
8.3.3 &  & $4 b_{\on}$ &  & $n\pi$ & $0$ & $0$ & $0$ & $0$ &  & $0$ & $1$ & $-\dfrac{\delta}{r_{\h}}$ & \\[11pt]
8.3.4 &  & $4 b_{\on}$ &  & $n\pi$ & $0$ & $0$ & $0$ & $0$ &  & $\dfrac{4\beta_{\on}}{r_{\h}}$ & $1$ & $\dfrac{\delta}{r_{\h}}$ & \\[11pt]
\hline
\end{tabular}
\caption{Parameter values that satisfy both the AFE and SFE for the case \( u = 0 \) and \( v\leq -1 \), where $\delta=\pm 1$.}
\label{FEcase8}
\end{table}

\section{Exact solutions}\label{app}

Finding exact vacuum solutions to the SFE \eqref{FESV}, while leveraging insights from the analysis of the AFE regarding the arbitrary functions \(\psi\) and \(\chi\), requires specific manipulations of \eqref{FESV}.
 In cases where exact solutions are feasible, such as those described in \cite{hayashi1967, obukhov672003}, we identify a suitable linear combination of the SFE that leads to integrable expressions. We then proceed with this preferred choice, given by
\begin{equation}
E_{1}=-(W_{\sl 11\sr}+W_{\sl 22\sr}+2W_{\sl 33\sr}) \quad \text{and} \quad E_{2}=-2(W_{\sl 22\sr}+W_{\sl 33\sr}) \, ,
\end{equation}
which, by utilizing Eqs.~\eqref{FESV} and \eqref{FV}, have the advantage of being expressible as:
\begin{subequations}\label{int1}
\begin{alignat}{1}\label{E1}
E_{1}: &\, \left[\left(A_{3}/A_{2}\right)^{2}G_{1}\right]'+\left(A_{3}/A_{2}\right)^{2}G_{1} [\ln( A_{1}A_{2})]'  =0 \, , \\[1.5ex]\label{E2}
E_{2}: & \,  \left[\left(A_{3}/A_{2}\right)^{3}G_{2}\right]'+\left(A_{3}/A_{2}\right)^{3}G_{2}\left([\ln (A_{1}A_{2}{}^{2}/A_{3})]'+A_{2}/A_{3}\right)=G_{3}\, ,
\end{alignat}
\end{subequations}
where
\begin{subequations}
\begin{alignat}{1}
&G_{1}=(b_{1}-b_{2})[\ln A_{1}]'+2b_{1}[\ln A_{3}]' +2b_{1}\frac{A_{2}}{A_{3}} \cos\chi\cosh\psi \, , \\[1.5ex]
&G_{2}=(2b_{1}-b_{2})[\ln A_{1}]'+(4b_{1}-b_{2})\left([\ln A_{3}]'-\frac{A_{2}}{A_{3}}\right)+2b_{1}\frac{A_{2}}{A_{3}}\left(1+\cos\chi\cosh\psi \right) \, , \\[1.5ex]\label{G3}
&G_{3}=2b_1 \frac{A_{3}}{A_{2}}\left(1+\frac{A_{3}}{A_{2}}[\ln A_{3}]'\right)\left(1+\cos\chi\cosh\psi \right)+2b_{3}\frac{A_{3}}{A_{2}}\left(2\sin^{2}\chi+\frac{A_{3}}{A_{2}}[\cos\chi]'\cosh\psi \right) \, .
\end{alignat}
\end{subequations}
Note that Eq.~\eqref{E1} can be integrated straightforwardly, while Eq.~\eqref{E2} cannot due to the presence of the term \(A_{2}/A_{3}\) and the dependencies of \(G_{3}\) on the arbitrary functions \(A_{2}\), \(A_{3}\), \(\psi\), and \(\chi\). Nevertheless, we rewrite Eqs.~\eqref{E1} and \eqref{E2} in the form:
\begin{equation}\label{FEG}
G_{1}=\frac{c_{1} A_{2}}{A_{1}A_{3}{}^{2}} \, , \quad G_{2}=\frac{\mu A_{2} }{A_{1}A_{3}{}^{2}}(c_{2}+G_{4}) \, ,
\end{equation}
with
\begin{equation}\label{mu}
\mu=\exp\left[-\int\frac{A_{2}}{A_{3}}\,dr \right] \, , \quad  G_{4}= \int\frac{A_{1}A_{2}{}^{2}}{\mu A_{3}}G_{3}\, dr  \, .
\end{equation}
An alternative and useful expression can be obtained by subtracting the equations in \eqref{FEG}, leading to
\begin{equation}\label{FEGP}
b_{1}[\ln A_{1}]'+(2b_{1}-b_{2})\left([\ln A_{3}]'-\frac{A_{2}}{A_{3}}\right) = \frac{\mu A_{2}}{A_{1}A_{3}{}^{2}}\left(c_{2}-c_{1}\mu^{-1}+G_{4} \right) \, .
\end{equation}
In this formulation, all dependencies on $\chi$ and $\psi$ are retained within the function $G_{4}$. Obtaining a second integration of Eq.~\eqref{E1} and performing the first integration of Eq.~\eqref{E2} will depend on specific choices of $\psi$ and $\chi$ that also satisfy the AFE \eqref{FEAV}. This task becomes particularly easier with a convenient choice of $\chi$ and $\psi$ that sets $G_{3} = 0$, along with an appropriate choice of coordinates that renders the integrand in Eq.~\eqref{mu} integrable, such as assuming $A_{3}(r') = f(r') A_{2}(r')$ for some given function $f(r')$.

\subsection{Choice of coordinates}

Given the static case, where all arbitrary functions are time-independent, we introduce new coordinates of the form \( t' = t \) and \( r' = A_{3}/A_{2} \) \cite{robert2024}. By adopting the so-called isotropic coordinates, this relation simplifies to
\begin{equation}\label{coor2}
A_{3} = r A_{2} \, .
\end{equation}
These coordinates render the spatial part of the metric conformally flat. In this setup, the coordinate \( r \) represents a radial distance in a spherically symmetric, isotropic 3-space, and is therefore generally restricted to \( r \geq 0 \).

This coordinate system was employed in~\cite{hayashi1967,obukhov672003} to obtain vacuum solutions in NGR, which is also the aim of this section. From this point onward, we will work with the isotropic coordinates defined in Eq.~\eqref{coor2}.

\subsection{AFE}

As previously discussed in Section~\ref{AFE}, when \(\chi = \chi_{\n}\) and \(\psi = \psi_{\n}\) are constants, the AFE undergo a significant simplification, as shown in Eq.~\eqref{AFEcp}. In the coordinates defined by Eq.~\eqref{coor2}, these equations take the form
\begin{subequations}\label{FEAVI}
\begin{alignat}{1} \label{FEAVIa}
& \cos\chi_{\n} \sinh\psi_{\n} = 0 \, , \\[2ex] \label{FEAVIb}
& \left(\frac{2 b_{3}}{r} \cos\chi_{\n} + F_{4} \cosh\psi_{\n}\right) \sin\chi_{\n} = 0 \, .
\end{alignat}
\end{subequations}
In order to satisfy Eq.~\eqref{FEAVI} while ensuring that \( \psi \) remains real, we set \( \psi_{\n} = 0 \) and \( \chi_{\n} = n\pi \), leading to \( \cos(n\pi) = \delta = \pm 1 \).
 However, this still leaves \(G_{4}\) with a complicated expression due to \(G_{3}\), given by
\begin{equation}
G_{3} = 2b_{1}r\left(1 + r[\ln(rA_{2})]'\right)(1+\delta) \, .
\end{equation}
A convenient choice is to set \(\delta = -1\), which requires \(n\) to be an odd integer. In this case, \(G_{3} = 0\), and \(G_{4}\) becomes a constant that can be absorbed into \(c_{2}\) in Eq.~\eqref{FEG}.

\subsection{SFE}

We now outline the procedure presented in~\citep{obukhov672003} for finding solutions to Eq.~\eqref{FEG}. By substituting \( \psi_{\n} = 0 \) and \( \chi_{\n} = n\pi \), where \( n \) is an odd integer, and adopting the coordinate choice from Eq.~\eqref{coor2}, which simplifies Eq.~\eqref{mu} to \( \mu = r^{-1} \), the system of equations in Eq.~\eqref{FEG} reduces to:
\begin{subequations} \label{SI2}
\begin{alignat}{1}
&(b_{1}-b_{2})[\ln A_{1}]'+2b_{1}[\ln A_{2}]'= \frac{c_{1} \, r^{-2}}{2A_{1}A_{2}} \,  ,\\[1.5ex]
&(2b_{1}-b_{2})[\ln A_{1}]'+(4b_{1}-b_{2})[\ln A_{2}]'=c_{2}\frac{r^{-3}}{A_{1}A_{2}} \, .
\end{alignat}
\end{subequations}
Eqs.~\eqref{SI2} constitute a system for $[\ln A_{1}]'$ and $[\ln A_{2}]'$ that can be algebraically solved to obtain
\begin{equation}\label{SRR1}
[\ln A_{1}]'=\frac{(b_{2}-4b_{1})c_{1}r^{-2}+4b_{1}c_{2}r^{-3}}{2(3b_{1}-b_{2})b_{2}\,A_{1}A_{2}} \quad \text{and} \quad [\ln A_{2}]'=\frac{(2b_{1}-b_{2})c_{1}r^{-2}-2(b_{1}-b_{2})c_{2}r^{-3}}{2(3b_{1}-b_{2})b_{2}\,A_{1}A_{2}} \, .
\end{equation} 
Adding these equations yields a direct solution for the function \( A_{1}A_{2} \), which can be written as
\begin{equation}\label{AA1}
A_{1}A_{2}=\frac{1}{b_{2}(3b_{1}-b_{2})r}\left(b_{1}c_{1}-(b_{1}+b_{2})\frac{c_{2}}{2r} \right)+c_{3} \, ,
\end{equation}
where \( c_{1} \), \( c_{2} \), and \( c_{3} \) are integration constants. Exploiting their arbitrariness, we can conveniently choose \( c_{3} \) as
\begin{equation}
c_{3}=-\frac{(4b_{1}-b_{2})c_{1}^{2}}{8b_{2}(3b_{1}-b_{2})c_{2}} \, .
\end{equation}
This choice of \( c_{3} \) allows us to rewrite Eq.~\eqref{AA1} in a more convenient form, given by
\begin{equation}\label{AA2}
A_{1}A_{2}=\gamma_{1}\gamma_{2}\left(\frac{c_{2}}{c_{1}r}-\frac{\alpha}{2}\right)\left(\frac{c_{2}}{c_{1}r}-\frac{\beta}{2}\right) \, ,
\end{equation}
where
\begin{equation}\label{gm1gm2}
\gamma_{1}\gamma_{2}=\frac{(b_{1}+b_{2})c_{1}^{2}}{2b_{2}(b_{2}-3b_{1})c_{2}} \, ,
\end{equation}
and the parameters \( \alpha \) and \( \beta \) are given by
\begin{equation}\label{alphabeta}
\alpha=\frac{2b_{1}-\sqrt{b_{2}(b_{2}-3b_{1})}}{b_{1}+b_{2}} \quad \text{and} \quad  \beta=\frac{2b_{1}+\sqrt{b_{2}(b_{2}-3b_{1})}}{b_{1}+b_{2}}\, .
\end{equation}
Note that Eq.~\eqref{gm1gm2} requires \( c_{2} \neq 0 \), \( b_{2} \neq 0 \), and \( b_{2} \neq 3b_{1} \). In addition, for the expressions in Eq.~\eqref{alphabeta} to yield real and physically meaningful values for \( \alpha \) and \( \beta \), we must also impose the conditions \( b_{2}(b_{2} - 3b_{1}) > 0 \) and \( b_{2} \neq -b_{1} \). Together, these constraints give rise to the following cases:
\begin{align}
&\begin{array}{lll}
\text{1)}\; b_{1} \leq 0 \text{ and } b_{2} < 3b_{1} 
& \quad \text{4)}\; b_{1} > 0 \text{ and } b_{2} < -b_{1} \\[0.3em]
\text{2)}\; b_{1} \leq 0 \text{ and } 0 < b_{2} < -b_{1} 
& \quad \text{5)}\; b_{1} > 0 \text{ and } -b_{1} < b_{2} < 0 \\[0.3em]
\text{3)}\; b_{1} \leq 0 \text{ and } b_{2} > -b_{1} 
& \quad \text{6)}\; b_{1} > 0 \text{ and } b_{2} > 3b_{1}
\end{array}
\label{casesb1b2}
\end{align}
Using the constraints in Eq.~\eqref{casesb1b2}, we determine the corresponding ranges of \( \alpha \) and \( \beta \), as given in Eq.~\eqref{alphabeta}, for each of the previously listed cases.
\begin{align}
&\begin{array}{lll}
\text{1)}\; \frac{1}{2}<\alpha\leq 1 \text{ and } -1 \leq \beta <\frac{1}{2} 
& \quad \text{4)}\; 1<\alpha<\frac{5}{4} \text{ and } \beta<-1 \\[0.3em]
\text{2)}\; \alpha >2 \text{ and } \frac{5}{4} < \beta <2 
& \quad \text{5)}\; \frac{5}{4}<\alpha<2 \text{ and } \beta >2 \\[0.3em]
\text{3)}\; \alpha \leq -1 \text{ and } 1\leq \beta <\frac{5}{4} 
& \quad \text{6)}\;-1<\alpha<\frac{1}{2} \text{ and } \frac{1}{2}<\beta <1
\end{array}
\label{casesalbe}
\end{align}
Furthermore, all these cases can be summarized by eliminating the square root in Eq.~\eqref{alphabeta} and expressing one parameter directly in terms of the other:
\begin{equation}\label{ab}
\alpha = \frac{4 - 5\beta}{5 - 4\beta}  \quad \text{or} \quad \beta = \frac{4 - 5\alpha}{5 - 4\alpha}
\end{equation}
This explicitly shows that \( \alpha \) and \( \beta \) are not independent parameters, as was already evident from the explicit expressions in Eq.~\eqref{alphabeta}, where both clearly depend on \( b_{1} \) and \( b_{2} \).

Having determined the permissible values for \( b_{1} \) and \( b_{2} \) in Eq.~\eqref{casesb1b2}, and consequently the corresponding values of \( \alpha \) and \( \beta \) in Eq.~\eqref{casesalbe}, we now use Eqs.~\eqref{AA1} and \eqref{AA2} to write the final solution in the form:
\begin{equation}\label{A1A2S}
A_{1} = \gamma_{1}\left(\frac{M}{r} - \frac{\alpha}{2}\right)^{\alpha}\left(\frac{M}{r} - \frac{\beta}{2}\right)^{\beta} 
\quad \text{and} \quad 
A_{2} = \gamma_{2}\left(\frac{M}{r} - \frac{\alpha}{2}\right)^{1 - \alpha}\left(\frac{M}{r} - \frac{\beta}{2}\right)^{1 - \beta} \, ,
\end{equation}
where \( M = c_{2}/c_{1} \), and \( \gamma_{1} \) and \( \gamma_{2} \) are related by Eq.~\eqref{gm1gm2}. These expressions for \( A_{1} \) and \( A_{2} \) are consistent with the results presented in~\citep{obukhov672003}. For the solution to be valid, it is essential that \( c_{1} \neq 0 \). 
Moreover, it is important to note that TEGR arises only in cases 1) and 3), and can be recovered by setting:
\begin{itemize}
    \item[i)] For case 1): $\alpha=1$ or \( b_{1} = 0 \), \( b_{2} < 0 \), and \( c_{1}{}^{2} = -8b_{2}c_{2} \) with \( c_{2} > 0 \).
    \item[ii)] For case 3): $\alpha=-1$ or  \( b_{1} = 0 \), \( b_{2} > 0 \), and \( c_{1}{}^{2} = -8b_{2}c_{2} \) with \( c_{2} < 0 \).
\end{itemize}
Using either of these options, one recovers the well-known TEGR solution in isotropic coordinates:
\begin{equation}\label{TEGRS}
A_{1} = \left(1 - \frac{2M}{r}\right)\left(1 + \frac{2M}{r}\right)^{-1} 
\quad \text{and} \quad 
A_{2} = \left(1 + \frac{2M}{r}\right)^{2} \, .
\end{equation}
In GR, this solution possesses an AH located at \( r_{s} = 2M \), which coincides with the event horizon. The spacetime also contains a curvature singularity at \( r = 0 \). We proceed to investigate the locations of the AH in the NGR solution given by Eq.~\eqref{A1A2S}, as detailed in the next section.

\subsection{AH}
To locate the AH in the spacetime described by Eq.~\eqref{A1A2S}, we compute \( \theta_{\sl \ell \sr} \) and \( \Delta \theta_{\sl \ell \sr} \), as detailed in Section~\ref{ENC}, using the coordinate choice specified in Eq.~\eqref{coor2} and the explicit expressions for \( A_{1} \) and \( A_{2} \) given in Eq.~\eqref{A1A2S}. After simplification, the results reduces to
\begin{subequations}
\begin{alignat}{1}\label{thl}
\theta_{\sl \ell \sr}=&\frac{\sqrt{2}}{4\gamma_{2}\,r^{3}}\left(\frac{M}{r}-\frac{\alpha}{2} \right)^{\alpha-2}\left(\frac{M}{r}-\frac{\beta}{2} \right)^{\beta-2} R_{1}\, , \\[10pt]\label{Dthl}
\Delta \theta_{\sl \ell \sr} = & \frac{1}{16\gamma_{2}{}^{2}\,r^{6}}\left(\frac{M}{r}-\frac{\alpha}{2} \right)^{2(\alpha-2)}\left(\frac{M}{r}-\frac{\beta}{2} \right)^{2(\beta-2)}R_{2},
\end{alignat}
\end{subequations}
where
\begin{subequations}
\begin{alignat}{1}
R_{1}=& \alpha\beta r^{2}-4M\alpha\beta r+4(\alpha+\beta-1)M^{2}\, , \\ 
R_{2}=& \alpha^{2} \beta^{2} r^{4}+ 2 M \alpha \beta (\alpha + \beta - 6 \alpha \beta) r^{3} + 8 M^{2} \alpha \beta (-3 + 2 \beta + 2 \alpha (1 + \beta)) r^{2}  \nonumber \\
&  - 8 M^{3} \left((-1 + \beta) \beta + \alpha^{2} (1 + 4 \beta) + \alpha (-1 + 4(-1 + \beta)\beta)\right) r +16 M^{4} (-1 + \alpha + \beta)^{2} \, . 
\end{alignat}
\end{subequations}
To verify the first condition in Eq.~\eqref{AHD2}, we set Eq.~\eqref{thl} equal to zero and find four distinct roots, each of which exists only for specific values of the parameters \( \alpha \) and \( \beta \). However, since all of them can be expressed in terms of constraints on \( \alpha \) via Eq.~\eqref{ab}, we rewrite them as follows:
\begin{subequations}\label{AHLv1}
\begin{alignat}{2}
r_{\alpha} &= 2M/\alpha, \quad &&\text{for } \alpha > 2, \\
r_{\beta}  &= 2M/\beta,  \quad &&\text{for } \frac{5}{4} < \alpha < 2, \\ 
r_{+}      &= 2M \left(1 + \sqrt{D} \right), \quad &&\text{for } \alpha \leq -1 \text{ or } 0 < \alpha < \frac{4}{5} \text{ or } \alpha \geq 1, \\
r_{-}      &= 2M \left(1 - \sqrt{D} \right), \quad &&\text{for } \alpha \leq -1 \text{ or } \alpha \geq 1,
\end{alignat}
\end{subequations}
where
\begin{equation}
D = \left(1 - 1/\alpha \right) \left(1 - 1/\beta \right) \, .
\end{equation}
Note that we have not considered the root \( r_{\h} = 0 \), as it is not physically meaningful in the context of spherically symmetric spacetimes expressed in isotropic coordinates~\eqref{coor2}. This is analogous to the exclusion of unphysical negative values \( r_{\h} < 0 \). 
Moreover, the roots \( r_{+} \) and \( r_{-} \) coincide at \( r_{s}=2M \) when \( \alpha = -1 \) or \( \alpha = 1 \), corresponding to \( D = 0 \) (TEGR).
By intersecting the intervals of permissible values for \( \alpha \) from the cases outlined in Eq.~\eqref{casesalbe} with those corresponding to the possible AH in \eqref{AHLv1}. The results are summarized in Table~\ref{Thv}.

\begin{table}[!h]
\centering
\renewcommand{\arraystretch}{1.4}
\begin{tabular}{|c|c|c|c|c|}
\hline
\textbf{Case} & \( \theta_{\sl \ell \sr}( r_{\alpha})=0 \) & \(  \theta_{\sl \ell \sr}(r_{\beta})=0 \) & \(  \theta_{\sl \ell \sr}(r_{+})=0 \) & \( \theta_{\sl \ell \sr}( r_{-})=0 \) \\
\hline
1) &   &   & \( \frac{1}{2} < \alpha < \frac{4}{5} \text{ or } \alpha = 1 \) & \( \alpha = 1 \) \\
\hline
2) & \( \alpha > 2 \) &   & \( \alpha > 2 \) & \( \alpha > 2 \) \\
\hline
3) &   &   & \( \alpha \leq -1 \) & \( \alpha \leq -1 \) \\
\hline
4) &   & \  & \( 1 < \alpha < \frac{5}{4} \) & \( 1 < \alpha < \frac{5}{4} \) \\
\hline
5) &   & \( \frac{5}{4} < \alpha < 2 \) & \( \frac{5}{4} < \alpha < 2 \) & \( \frac{5}{4} < \alpha < 2 \) \\
\hline
6) &   &   & \( 0 < \alpha < \frac{1}{2} \) &   \\
\hline
\end{tabular}
\caption{Parameter ranges of \( \alpha \) for which \( \theta_{\sl \ell \sr}(r_{\h}) = 0 \) holds in each case; a blank indicates that no such range exists.
}\label{Thv}
\end{table}

To verify the second condition in Eq.~\eqref{AHD2}, we evaluate Eq.~\eqref{Dthl} at \( r < r_{\h} \), corresponding to a region in the neighbourhood just inside the possible AH. We then check whether the result is a real and negative number, which would indicate the existence of a well-defined AH. The corresponding results are summarized in Table~\ref{DThv}.

\begin{table}[!h]
\centering
\renewcommand{\arraystretch}{1.4}
\begin{tabular}{|c|c|c|c|c|}
\hline
\textbf{Case} & \( \Delta \theta_{\sl \ell \sr}(r< r_{\alpha})<0 \) & \( \Delta \theta_{\sl \ell \sr}(r< r_{\beta})<0 \) & \( \Delta \theta_{\sl \ell \sr}(r< r_{+})<0 \) & \( \Delta \theta_{\sl \ell \sr}(r< r_{-})<0 \) \\
\hline
1) &   &   & \(\alpha = 1 \) & \( \alpha = 1 \) \\
\hline
2) &  &   &  &   \\
\hline
3) &   &  & \( \alpha = -1 \) & \( \alpha \leq -1 \) \\
\hline
4) &   &   &  & \( 1 < \alpha < \frac{5}{4} \) \\
\hline
5) &   &    & \(  \alpha =\frac{3}{2} \) & \( \alpha = \frac{7}{5} \) \\
\hline
6) &  &   &   &  \\
\hline
\end{tabular}
\caption{Parameter values of \( \alpha \) for which \( \Delta \theta_{\sl \ell \sr}(r < r_{\h}) < 0 \) holds in each case; a blank indicates that no such value exists.
}\label{DThv}
\end{table}

Table~\ref{DThv} shows that cases 2) and 6) do not possess any AH. Furthermore, \( r_{\alpha} \) and \( r_{\beta} \) do not define an AH in any of the cases or for any admissible values of the parameter \( \alpha \). This result stands in stark contrast to the assumption made in~\citep{obukhov672003}, where \( r_{\alpha} \) and \( r_{\beta} \) were taken to represent the horizon locations for the solution given in Eq.~\eqref{A1A2S}. That conclusion arises from a coordinate-dependent approach, in which the horizon is identified by the vanishing of the temporal component of the metric, specifically, the condition \( -A_{1}^2(r_{\h}) = 0 \). While this criterion can be useful in certain coordinate systems, it is only reliable in special cases where it coincides with the geometric prediction based on the vanishing of the expansion scalars.

\begin{table}[!h]
\centering
\renewcommand{\arraystretch}{1.4}
\begin{tabular}{|c|c|c|}
\hline
\textbf{Case}  &  $r_{+}$ & $ r_{-}$ \\
\hline
1) & \(\alpha = 1 \) & \( \alpha = 1 \) \\
\hline
3.1) & \( \alpha = -1 \) & \( \alpha = -1 \) \\
\hline
3.2) &  & \( \alpha < -1 \) \\
\hline
4) &   & \( 1 < \alpha < \frac{5}{4} \) \\
\hline
5) & \(  \alpha =\frac{3}{2} \) & \( \alpha = \frac{7}{5} \) \\
\hline
\end{tabular}
\caption{Parameter values of \( \alpha \) for which a well-defined AH exists in each case; a blank indicates that no such value exists.
}\label{rprn}
\end{table}
\newpage 
Table~\ref{rprn} presents the values of \( \alpha \) for which an AH is well defined. Note that case 3) has been split into two subcases: 3.1) and 3.2). Cases 1) and 3.1) both correspond to TEGR, differing only in the values of \( b_{2} \) and \( c_{2} \), as previously discussed in the derivation of Eq.~\eqref{TEGRS}. Since they ultimately describe the same solution, we discard case 3.1) and retain the remaining cases for further analysis.

\subsection{Teleparallel scalars analysis}
Using the expressions in Eqs.~\eqref{A1A2S} and \eqref{TS1}, we compute the teleparallel scalars listed in Eq.~\eqref{TScalars} for the coordinate system defined in Eq.~\eqref{coor2}, and for the values \( \chi_{\n} = n\pi \) and \( \psi_{\n} = 0 \), with \( n \) an odd integer. This yields:
\begin{subequations}\label{TSSI}
\begin{alignat}{1}
\mathscr{V}=&\,\frac{
M^2 R_{v}{}^{2} }{
\gamma_{2}{}^2\, r^6}\left( \frac{M}{r}-\frac{\alpha}{2}  \right)^{2(\alpha-2)} 
\left(\frac{M}{r}-\frac{\beta}{2} \right)^{2(\beta-2)}, \\[7pt]
\mathscr{A}=&\, 0 ,\\
\mathscr{T}=&\,\frac{
M^2 R_{t}{}^{2} }{
4\gamma_{2}{}^2\, r^6}\left( \frac{M}{r}-\frac{\alpha}{2}  \right)^{2(\alpha-2)} 
\left(\frac{M}{r}-\frac{\beta}{2} \right)^{2(\beta-2)}, \\[7pt]
T=&\,\frac{M^{2}R_{+}R_{-}}{2\gamma_{2}{}^{2}\, r^{6}}\left(\frac{M}{r}-\frac{\alpha}{2}\right)^{2(\alpha-2)}\left(\frac{M}{r}-\frac{\beta}{2}\right)^{2(\beta-2)} \, ,
\end{alignat}
\end{subequations}
where
\begin{subequations}
\begin{alignat}{1}
R_{v}=&\, M(\alpha + \beta-4 ) + (\alpha + \beta - \alpha\beta) r\, ,\\
R_{t}=& \,4M(\alpha + \beta -1 ) + (\alpha + \beta - 4\alpha\beta)r\, ,  \\
R_{\pm}=&\,2(\alpha+\beta\pm 2)m\mp (\alpha+\beta\pm 2\alpha\beta)r \, .
\end{alignat}
\end{subequations}
With the exception of \( \mathscr{A} \), all of these scalars diverge at \( r = 0 \) for any admissible value of \( \alpha \) in the cases where an AH exists. We now evaluate the scalars at the AH locations listed in Table~\ref{rprn}, excluding case 3.1), to assess whether the region \( r = r_{\h} \) and its interior are part of the spacetime manifold. The results are summarized in Table~\ref{TSAH}.

\begin{table}[!h]
\centering
\renewcommand{\arraystretch}{1.4}
\begin{tabular}{|c|c|c|c|c|c|c|}
\hline
\textbf{Case} & \( \mathscr{V}(r_{+}) \) & \( \mathscr{T}(r_{+}) \) & \( T(r_{+}) \) & \( \mathscr{V}(r_{-}) \) & \( \mathscr{T}(r_{-}) \) & \( T(r_{-}) \) \\
\hline
1)   & \( \infty \) & \( \infty \) & \( \infty \) & \( \infty \) & \( \infty \) & \( \infty \) \\
\hline
3.2) &  &   &  & \( > 0 \) & \( > 0 \) & \( < 0 \) \\
\hline
4)   &  &  &  & \( > 0 \) & \( > 0 \) & \( < 0 \) \\
\hline
5)   & \( > 0 \) & \( > 0 \) & \( > 0 \) & \( > 0 \) & \( > 0 \) & \( < 0 \) \\
\hline
\end{tabular}
\caption{Torsion scalars evaluated at the AH, \( r_{+} \) and \( r_{-} \), for each case; a blank indicates that no AH exists to be evaluated in the corresponding case.
}
\label{TSAH}
\end{table}

As shown in Table~\ref{TSAH}, for case 1) we have \( r_{+} = r_{-} = r_{s} \). Notably, case 1) is the only case in which the teleparallel scalars diverge at the apparent horizon location. It is also the only model among those considered that possesses a proper Newtonian limit. This model corresponds to Type 2 (or Type 6 if \( b_{3} = 0 \)) in Table~\ref{GFC}, and is ghost-free provided that \( b_{2} > 0 \).

In contrast, case 3.2) features convergent scalars and can be classified as Type 3 if \( b_{3} > 0 \) and \( b_{1} = -b_{3} \). However, this represents a theory that does not recover TEGR in any limit. Similarly, cases 4) and 5) fall under Type 1, which is characterized by unavoidable ghost modes and, again, does not admit TEGR as a limiting case, since in these cases \( b_{1} > 0 \).

These results are consistent with the broader analysis presented in the main text and serve as a concrete example evaluated in a different coordinate gauge.

\end{document}